\documentclass[journal,10pt,a4paper,twocolumn,twoside]{IEEEtran}

\IEEEoverridecommandlockouts

\usepackage{cite}
\usepackage{amsmath,amssymb,amsfonts}
\usepackage{amsthm}
\usepackage{tikz}
\usepackage{graphicx}
\usepackage{textcomp}
\usepackage{xcolor}
\usepackage{caption} 
\usepackage{subcaption}
\usepackage{hyperref}
\usepackage{mathtools}
\usepackage{algorithm}
\usepackage{algpseudocode}
\usepackage{booktabs}
\usepackage{longtable,multicol}
\usetikzlibrary{automata, positioning, arrows,calc}
\usetikzlibrary{quantikz}
\usepackage{tcolorbox}
\usepackage{cuted}
\usepackage{multicol}
\usepackage{multirow}

\theoremstyle{plain}
\newtheorem{property}{Property}
\newtheorem*{definition*}{Definition}
\newtheorem{definition}{Definition}
\newtheorem{theorem}{Theorem}
\newtheorem{lemma}{Lemma}
\newtheorem{prop}{Proposition}
\newtheorem*{proof*}{Proof}

\theoremstyle{definition}

\newtheorem*{remark*}{Remark}
\newtheorem*{decfor}{Decision Formulation}
\newtheorem*{expectedRewards}{Expected Rewards}
\newtheorem*{whenToStop}{(Optimally) Knowing When to Stop}
\newtheorem*{rewardMM}{Reward Majorant and Minorant}

\newcommand{\eqdef}{\stackrel{\triangle}{=}}

\newsavebox\ltmcbox

\begin{document}

\title{Entanglement Distribution in the Quantum Internet: \textit{Knowing when to Stop!}}
\author{Angela~Sara~Cacciapuoti$^*$,~\IEEEmembership{Senior~Member,~IEEE}, Jessica Illiano, Michele Viscardi, Marcello~Caleffi,~\IEEEmembership{Senior~Member,~IEEE}
    \thanks{The authors are with the \href{www.quantuminternet.it}{www.QuantumInternet.it} research group, \textit{FLY: Future Communications Laboratory}, University of Naples Federico II, Naples, 80125 Italy. A.S. Cacciapuoti and M. Caleffi are also with the Laboratorio Nazionale di Comunicazioni Multimediali, National Inter-University Consortium for Telecommunications (CNIT), Naples, 80126, Italy. }
    \thanks{$^*$Corresponding author.}
    \thanks{A preliminary version of this work is currently under review for IEEE QCE23 \cite{VisIllCac-23}.}
    \thanks{Angela Sara Cacciapuoti acknowledges PNRR MUR NQSTI-PE00000023, Marcello Caleffi acknowledges PNRR MUR project RESTART-PE00000001.}
}

\maketitle

\begin{abstract}
Entanglement distribution is a key functionality of the Quantum Internet. However, quantum entanglement is very fragile, easily degraded by decoherence, which strictly constraints the time horizon within the distribution has to be completed. This, coupled with the quantum noise irremediably impinging on the channels utilized for entanglement distribution, may imply the need to attempt the distribution process multiple times before the targeted network nodes successfully share the desired entangled state. And there is no guarantee that this is accomplished within the time horizon dictated by the coherence times. As a consequence, in noisy scenarios requiring multiple distribution attempts, it may be convenient to stop the distribution process early. In this paper, we take steps in the direction of \textit{knowing when to stop} the entanglement distribution by developing a theoretical framework, able to capture the quantum noise effects. Specifically, we first prove that the entanglement distribution process can be modeled as a Markov decision process. Then, we prove that the optimal decision policy exhibits attractive features, which we exploit to reduce the computational complexity. The developed framework provides quantum network designers with flexible tools to optimally engineer the design parameters of the entanglement distribution process.
\end{abstract}


\section{Introduction}
\label{sec:introduction}
The Quantum Internet is foreseen to enable several applications with no counterpart in the classical world \cite{IllCalMan-22,RamPirDur-21,VanSatBen-21,CacCalTaf-20,WehElkHan-18,DurLamHeu-17,Kim-08}, such as distributed quantum computing \cite{CalAmoFer-22} and secure communications \cite{WanRahLi-22}. To this aim, the entanglement distribution process plays the \textit{key} role.  Indeed, the successful distribution of entangled states among remote network nodes represents a necessary condition for any entanglement-based network \cite{KozWehVan-23}.
\begin{figure*}
    \centering
    \resizebox{0.8\textwidth}{!}{
    \input{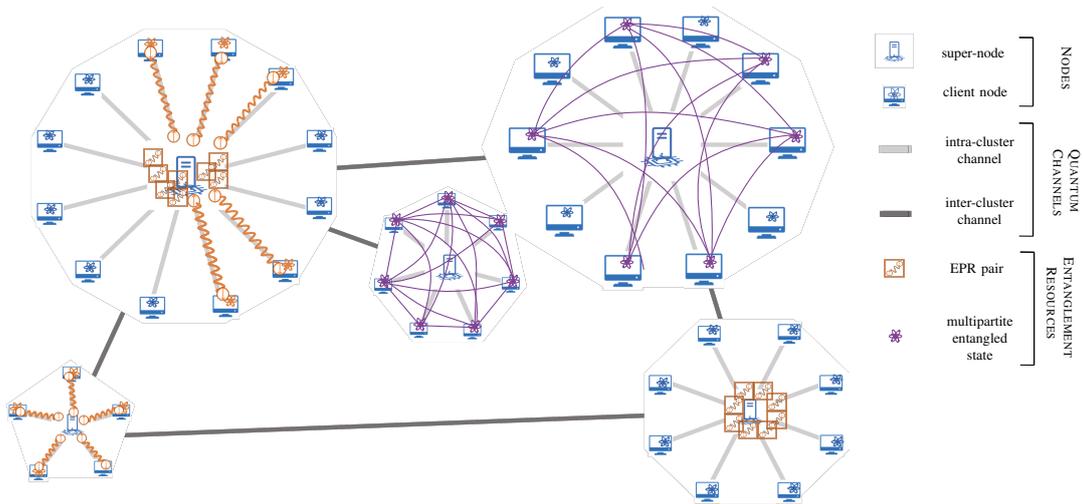}
    }
    \caption{Pictorial representation of the considered quantum network architecture. The quantum network comprises several clusters of nodes. Within each cluster, client nodes are connected to a super-node. The super-nodes are specialized nodes equipped with dedicated hardware able to generate the entanglement resources. Client nodes obtain access to the multipartite entangled states, generated at the super-nodes, through the entanglement distribution process.}
    \label{fig:1}
    \hrulefill
\end{figure*}
\begin{figure*}
    \centering
    \begin{subfigure}{0.5\textwidth}
        \centering
        \includegraphics[width=0.9\textwidth]{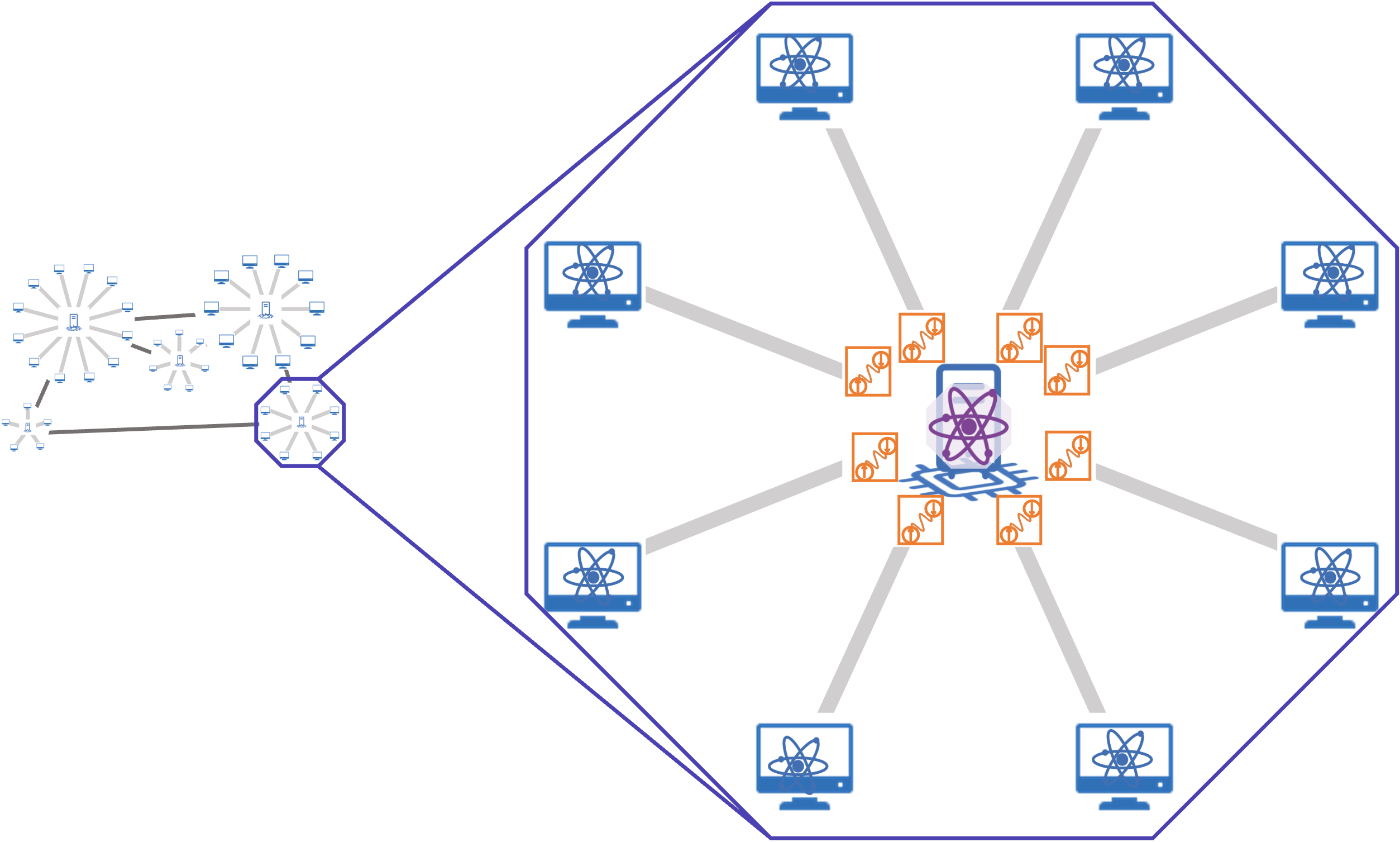}
        \subcaption{The super-node generates the EPR pairs (orange squares) and the multipartite entangled state (purple atom). }
        \label{fig:01a}
    \end{subfigure}
    \begin{subfigure}{0.45\textwidth}
        \centering
        \includegraphics[width=0.66\textwidth]{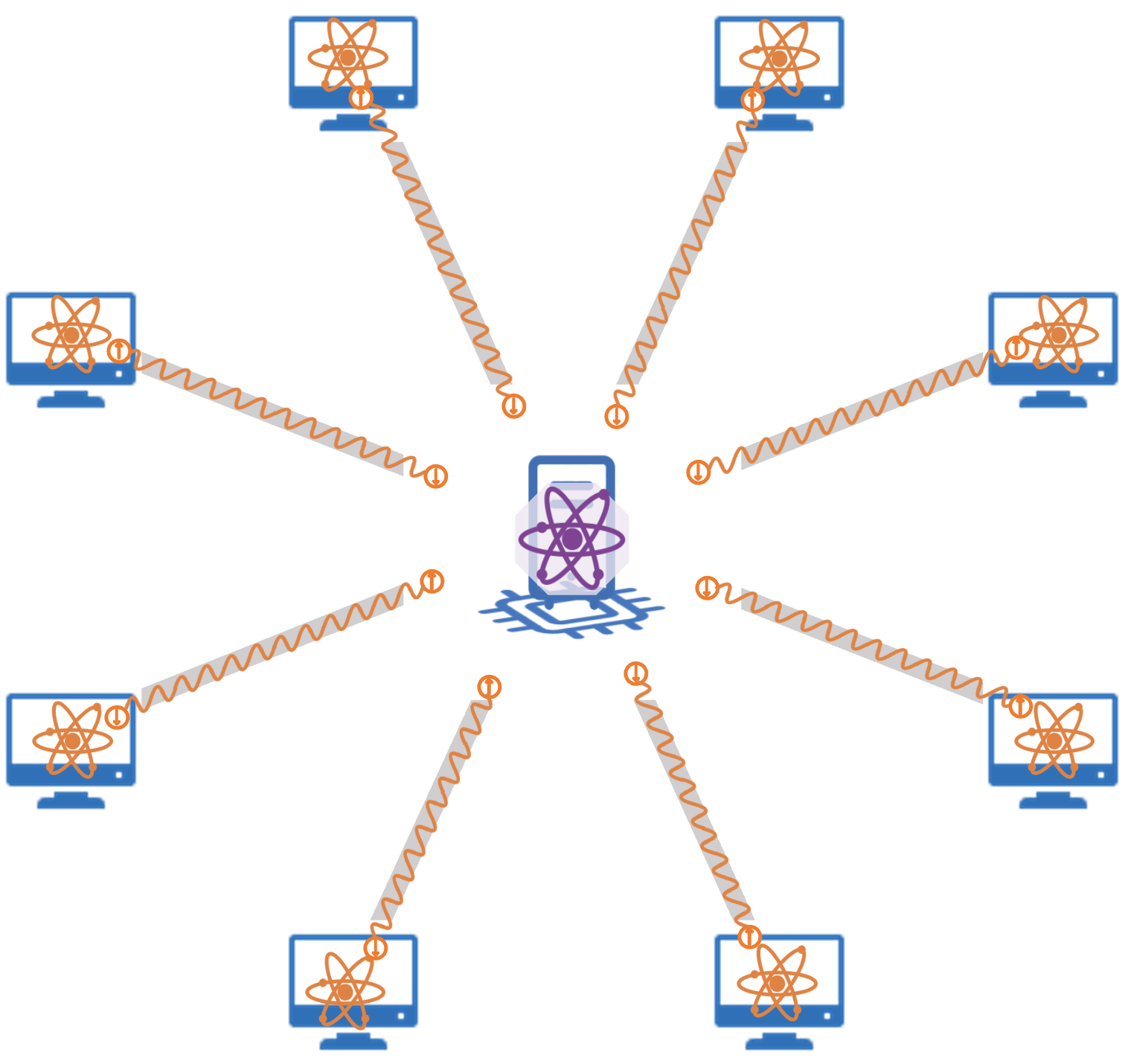}
        \subcaption{The super-node distributes the EPR pairs. One ebit is transmitted through the physical quantum channel (gray line).}
        \label{fig:01b}
    \end{subfigure}
    \begin{subfigure}{0.45\textwidth}
        \centering
        \includegraphics[width=0.66\textwidth]{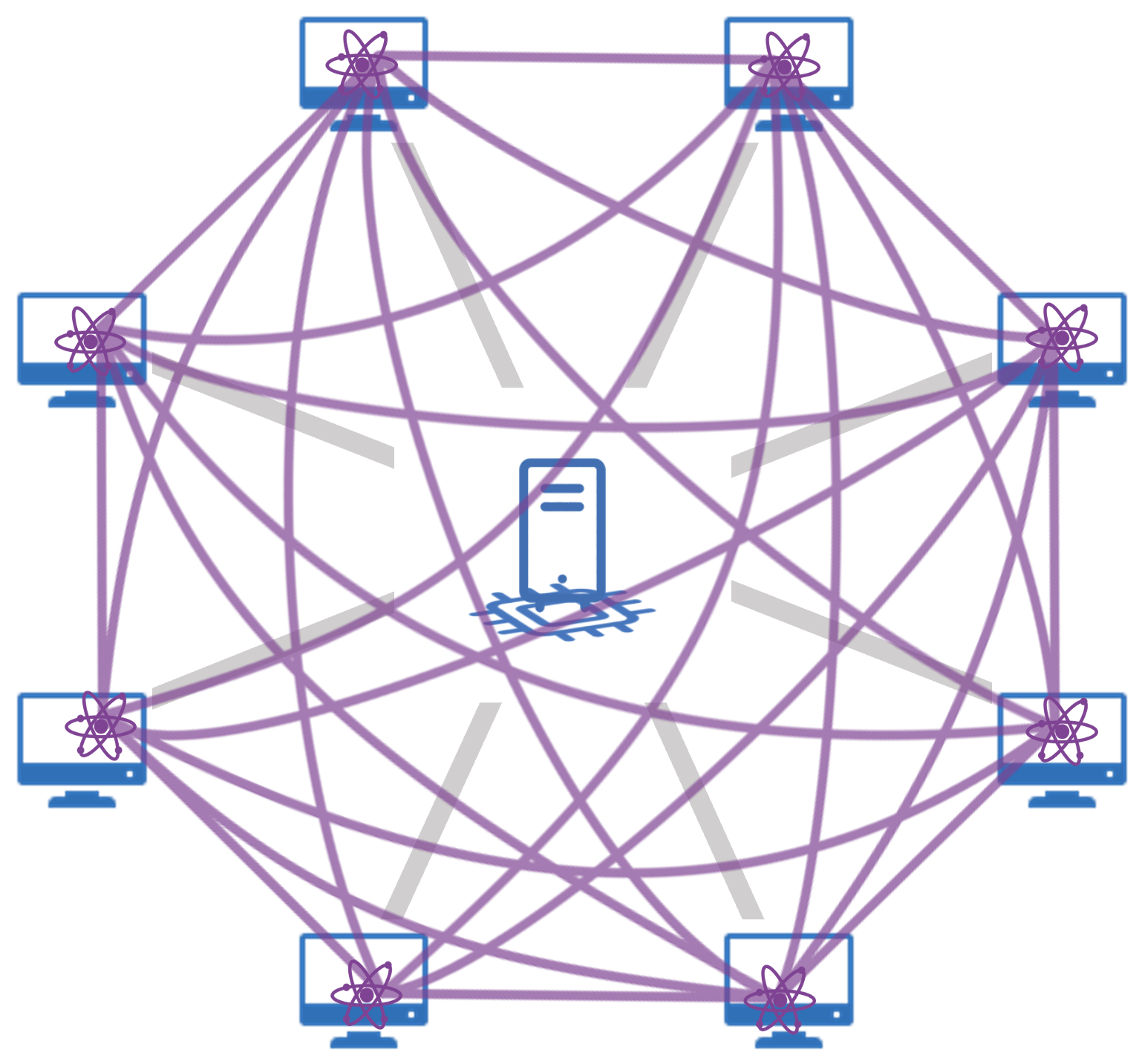}
        \subcaption{By exploiting the distributed EPR pairs, the multipartite entangled state is teleported to the client nodes.}
        \label{fig:01c}
    \end{subfigure}
    \caption{Pictorial representation of the system model. The legend of the figure available in Fig.~\ref{fig:1}. Subfigure (a) represents a zoomed-view of the cluster of nodes in the lower-right part of Fig.~\ref{fig:1}.  We consider a scenario where a super-node is connected through quantum channels to a set of quantum nodes, referred to as \textit{clients}. The super-node is in charge of generating and distributing EPR pairs to the clients. The aim of the process is to distribute the multipartite entangled state through teleportation. For this, the super-node performs multiple attempts of ebit  distribution, as represented in Fig.~\ref{fig:01b}. Fig.~\ref{fig:01b} constitutes the ideal scenario with the successful distribution of all the EPR pairs between the super-node and the clients, but clearly multiple distribution attempts might be required depending on the noise level affecting the quantum channels. Finally, the super-node can exploit the distributed EPR pairs for teleporting the multipartite entangled state. As represented in Fig.~\ref{fig:01c}, after the teleportation, the client nodes share a multipartite entangled state. }
    \label{fig:01r}
    \hrulefill
\end{figure*}
From a network design perspective, there exist two different approaches for entanglement generation and distribution: \textit{proactive} or \textit{reactive} strategies. Proactive strategies aim at early distribution of entanglement resources, where a new generation process ideally starts as soon as the entanglement resource is depleted. Differently, reactive strategies aim at on-demand distribution of entanglement, with a new generation process starting according to an entangled resource demand, namely, when needed \cite{IllCalMan-22}. With this distinction in mind, a few theoretical models and analysis of entanglement distribution have been recently proposed in literature \cite{LiXueLi-22, VanShc-22,DupGouGui-23,HalBarKha-23,VarGuhNai-21-1,VarGuhNai-21-2,IneVarSca-23}. In \cite{VarGuhNai-21-1}, the authors model the distribution of entangled pairs as a discrete time Markov chain. Specifically, they assume infinite coherence time and infinite resources at the central node, with the aim of analyzing the expected capacity of the central node in terms of the number of qubits to be stored to meet the stability condition of the system. In \cite{VarGuhNai-21-2}, the distribution of entangled pairs is modeled as a continuous time Markov chain. Such a model is based on a Poisson probability distribution for the successful distribution of entangled pairs, and it accounts for some non-idealities, such as decoherence and noisy measurements. In \cite{IneVarSca-23}, a Markov decision process is used to study the limits of bipartite entanglement distribution via entanglement swapping, by using a chain of quantum repeaters equipped with quantum memories. Finally, in \cite{KhaMatSid-19,Kha-22,Kha-21} some practical figures of merit for entanglement distribution in quantum repeater networks are provided. In particular, the authors define the average connection time and the average size of the largest distributed entangled state for a fixed scenario. 

Despite these research efforts, the fundamental problem of \textit{knowing when to stop} (the entanglement distribution) remains unsolved. And filling this research gap is mandatory for the efficient engineering of any entanglement distribution process. 

Specifically, it is well-known that quantum entanglement is a very fragile resource, easily degraded by decoherence \cite{CalCac-20,CacCalVan-20,WeiVanCoo-23}. Decoherence severely impacts the time horizon in which freshly-generated entangled states can be successfully distributed and exploited for communication needs. Yet, due to the noise irremediably affecting the quantum communication channels utilized for entanglement distribution, it may be necessary to attempt the distribution process multiple times before that all the selected network nodes successfully share the targeted entangled state.

As a matter of fact, because of the complex and stochastic nature of the physical mechanisms underlying quantum noise, there is no guarantee that all the selected nodes can successfully share the entangled state within the time horizon dictated by the coherence times. As a consequence, in noisy scenarios requiring multiple distribution attempts, it may be convenient to stop the distribution process early, i.e., before entangling all the selected nodes. The rationale for this choice is twofold. On one hand, an early stopping can be required to account for additional delays induced by the network functionalities exploiting the entanglement resource. On the other hand, an early stopping can be convenient whenever ``\textit{enough}'' nodes -- accordingly to a certain figure of merit -- already share entanglement, so that the entangled resource can be promptly exploited for the needed communication/computing purpose. 

In this paper, we take steps in the direction of \textit{knowing when to stop} by developing a theoretical framework. This framework provides quantum network designers with flexible tools to optimally engineer the design parameters of the entanglement distribution. To the best of our knowledge, this is the first work addressing the optimal stopping rule for entanglement distribution.

\subsection{Our contributions}
The developed theoretical framework abstracts from the particular state to be distributed and provides a model that can be tweaked to account for the physical characteristics of the process itself. Specifically through the paper:
\begin{enumerate}
    \item[-] we provide a comprehensive characterization of the entanglement distribution problem, by showing that it can be modeled as a Markov decision process with minimal assumptions;
    \item[-] we provide the optimality conditions of the policy to be adopted, and we prove some key properties of the optimal policy that can be exploited for reducing the computational complexity;
    \item[-] we analyze the impact of different reward functions on the distribution process through two main figures of merit: the average cluster size and the average distribution time;
    \item[-] we gain insights on the selection of appropriate reward functions for entanglement distribution process engineering.
\end{enumerate}

In summary, we present an easy-to-use tool for modeling and fine-tuning entanglement distribution systems to meet specific performance requirements. It is important to emphasize that the model we offer in this study is highly adaptable and can be tailored to various scenarios and applications.

The rest of the manuscript is organized as follows. In Sec.~\ref{sec:2}, we introduce the system model along with some preliminaries. In Sec.~\ref{sec:3}, we first formulate the entanglement distribution as a decision process, and then we derive both general (Sec.~\ref{sec:3.2}) and reward-dependent (Sec.~\ref{sec:3.3}) properties of the optimal policy, which we exploit for for reducing the computational complexity of the optimal policy search. In Sec.~\ref{sec:4} we validate the theoretical analysis through numerical simulations, and we discuss the impact of the reward functions on the performance of the entanglement distribution process. Finally, in Sec.~\ref{sec:5} we conclude the paper, and some proofs are gathered in the Appendix.

\section{System Model}
\label{sec:2}

Generating and distributing entanglement can be a demanding task due to the delicate nature of quantum states and their susceptibility to environmental disturbances. The complexity of entanglement generation and distribution becomes more evident for multipartite entangled states. Indeed, in many practical scenarios the generation of multipartite entanglement requires sophisticated and resource-intensive setups, often involving complex experimental apparatuses and precise control mechanisms. 
These technological limitations, coupled with the need for specialized environments that can facilitate quantum communication processes, make it pragmatic to assume a specialized super-node responsible for entanglement generation and distribution \cite{EppKamBru-17,AviRozWeh-22,VarGuhNai-21-1,IllCalVis-23}. Hence, it is reasonable to conceptualize an entanglement-based network architecture as the collection of clusters as represented in Fig.~\ref{fig:1}. In each cluster of quantum nodes, a super-node is connected through quantum channels to the other nodes, referred to as \textit{clients} in the following. In this context, only super-nodes are equipped with the aforementioned advanced apparatus and, hence, able to generate entanglement. As a consequence, in order obtain an entangled state shared between the client nodes, first the super-nodes should locally generate the entangled resource. Then, the entangled qubits (ebits) should be distributed to the clients according to a certain distribution strategy. 

 \begin{remark*}
 We emphasize that, when it comes to the distribution of multipartite entanglement, the assumption of a super-node for the entanglement generation is needed, not only due to the current maturity of the quantum technologies, but also due to the unavoidable requirement of some sort of local interaction among the qubits to be entangled, as discussed in \cite{IllCalVis-23}.
 \end{remark*}

\renewcommand{\arraystretch}{1.5}

\begin{table*}
    \centering
    \begin{subtable}[t]{0.45\textwidth}
    \centering
        
        \begin{tabular}[t]{  c | p{6cm}  }
        \hline

        \hline
        \multicolumn{1}{c}{Symbols} & {Definitions} \\
        \hline
            $p$ & probability of an ebit propagating through a quantum channel without experiencing absorption\\
        
            $q \eqdef 1-p$ & probability of ebit distribution failure as a consequence of the carrier absorption\\
            
            $F$& \textit{fidelity} of an EPR pair shared between client node and super-node\\
            $F_{th}$& \textit{fidelity theshold}: fidelity value of an EPR pair that can be considered as exploitable\\    
            
            $\mathcal{N}$ & \textit{time horizon}: set constituted by $N$ time slot\\
        
            $N$ & number of time slots within the coherence time \\

            $\mathcal{A}$ & \textit{action set}: set of actions available at the super-node\\

            $C$ & action of attempting another distribution round in the next time slot\\

            $Q$ & action of stopping (i.e., not attempting) the distribution\\

            $\mathcal{S}$ & set of possible values for the number of connected clients\\

            $\Delta$ & \textit{absorbing state}: state of the system where no further distributions are attempted.\\

            $\tilde{\mathcal{S}}$ & set of possible values for the state of the client set at any time slot\\

            $s\in \tilde{\mathcal{S}} $ & state of the client set \\

            $n\in \mathcal{N}$ & n-th time slot within the time horizon of the system\\
        
            $(s,n)\eqdef s_n $ & \textit{system state}: pair of client set state and considered time slot\\
        
            $\mathcal{A}_{s_n}$ & \textit{allowed action set}: set of action available at the super-node when the system state is $s_n$\\
        
            $r(s_n,a)$ & \textit{reward function}: overall reward achieved when the system is in state $s_n$ and action $a$ is taken\\
        
            $f(s_n)$ & \textit{continuation cost function}: function modelling the overall cost of further attempting the ebit distribution when the system is in $s_n$\\
        
            $g(s_n)$ & \textit{pay-off function}: function modelling the gain achievable by stopping the ebit distribution when the system is in $s_n$\\
            
        \end{tabular}
    \end{subtable}
    \begin{subtable}[t]{0.45\textwidth}
    \centering
        
        \begin{tabular}[t]{  c | p{6cm}  }
        \hline

        \hline
        \multicolumn{1}{c}{Symbols} & {Definitions} \\
        \hline
        
        $p(\tilde{s}_{n+1}| s_n)$ & \textit{transition probability}: probability of the system evolving into the state $\tilde{s}_{n+1}$ from the state $s_n$ when action $a$ is taken \\
        
        $p(\tilde{s}| s)$ & probability of having $\tilde{s}$ connected client nodes with one distribution attempt given that $s$ client node have already successfully received an ebit \\
        
        $\pi(\cdot)$ & \textit{policy}: rule determining the action to be taken in any possible state of the considered system.\\
        $v_\pi(s_1)$ & \textit{total expected reward}, recursively obtained starting from the initial state $s_1$\\
        $v_\pi(\tilde{s}_n)$ & \textit{expected remaining reward} at the timeslot $n$\\
        $\pi^*(s_1)$ & strategy that maximizes the expected total reward\\
        $p(s)$& probability of successfully distributing ebits to $s$ clients\\
        $v^*(s_1)\eqdef v_{\pi^*}(s_1)$& \textit{maximum expected total reward}: expected total reward achieved by the optimal policy\\
        $v^*_Q$& maximum expected reward achievable when action $Q$ is taken\\
        $v^*_C$& maximum expected reward achievable when action $C$ is taken\\
        $p(\breve{s}_{n+k}|s_n,C)$& \textit{extended transition probability}, probability to evolve into state $\breve{s}_{n+k}= (\breve{s}, n+k)$ at time slot $n+k$, starting from state $s_n=(s,n)$ with $s \neq \Delta$, by having chosen always action $C$ at the end of each of time-slot between $n$ and $n+k-1$\\
        
        $v^+(s_n)$& \textit{reward majorant}\\
        $v^-(s_n)$& \textit{reward minorant}\\
        $\mathcal{S}^Q_n$& \textit{OLA} set: one step look ahead set of system states where the instantaneous reward achievable by stopping is not lower than the expected reward achievable by attempting a further distribution attempt and then deciding to stop the distribution\\
        $\mathfrak{S}_{n+1}$& random variable describing the system state at step $n+1$\\
        
    \end{tabular}
    \end{subtable}
    
    \caption{Adopted Notation}
    \label{tab:01}
    \hrulefill
\end{table*} 
Accordingly, we consider a scenario where a super-node is in charge of generating and distributing EPR pairs to a set of $S$ clients through independent quantum channels. Thus, we assume each client holding at least one specialized qubit -- aka communication qubit \cite{KozWehVan-23,CalAmoFer-22} -- reserved for communication purposes. Similarly, we assume the super-node holding at least $S$ communication qubits to this purpose. It is worthwhile to note that the assumption of EPR pair distribution to client nodes is not restrictive, i.e., it does not hold only in EPR-based networks. Differently, the EPR distribution represents a key step for the distribution multipartite entangled states as well. In fact, when it comes to the distribution of multipartite entanglement, the super-node can, in principle, directly distribute each ebit to each client. However, this approach is not viable for all the classes of multipartite entanglement, which are characterized by different\footnote{As an example, the direct distribution of GHZ-like states, which are characterized by the lowest persistence, requires all the photons encoding the GHZ state to be successfully distributed to the clients in a single distribution attempt \cite{ZhoLiXia-23}.} \textit{persistence} properties\cite{IllCalMan-22}. Accordingly, in the following we consider the more general case in which multipartite entangled states are distributed through teleportation \cite{BugCouOma-23}, by exploiting the a-priori distribution of EPR pairs via heralded schemes \cite{BarCroZei-10,HofHruOrt-12}. As a matter of fact, this strategy is very common in literature and it has been proved also to guarantee more resilience to noise and better protection against memory decoherence \cite{AviRozWeh-22}. 

By accounting for the above, the considered system model is depicted in Fig.\ref{fig:01r}. Specifically, Fig.\ref{fig:01a} provides a zoomed-view of the cluster depicted in the lower-right section of Fig.~\ref{fig:1}. Within this framework, the super-node ultimate goal is to distribute a multipartite entangled state to the client nodes. To achieve this, the super-node locally generates two distinct entangled resources: one being the multipartite entangled state, and the other a set of EPR pairs necessary for teleporting the multipartite state.

Ideally, the super-node aims at achieving the scenario illustrated in Fig.~\ref{fig:01b}, wherein each client successfully receives an entanglement bit (ebit) corresponding to an EPR pair. However, the effects of noise imposes multiple attempts of ebit distribution to attain this scenario. Eventually, when the distribution of EPR pairs is terminated, the super-node proceeds to distribute the multipartite entangled state through teleportation, and, finally, as depicted in Fig.~\ref{fig:01c}, once teleportation has been performed, the clients collectively share a multipartite entangled state.

In the following we collect some definitions and assumptions that will be used in the paper and summarize the definitions as well as the notations used in Table~\ref{tab:01}.

\textbf{EPR Distribution Model}: The distribution attempt of an EPR ebit toward a client node through a noisy quantum channel is modeled with a Bernoulli distribution with parameter $p$, where $p$ denotes the successful distribution probability.

According to the above, we consider quantum channels modeled as absorbing channels. Such a model constitutes  a \textit{worst-case} scenario, since the noise irreversibly corrupts the information carrier without any possibility of ebit recovery \cite{BenDiVSmo-97,BenShoSmoTha-99,BruFaoMac-00,NieChu-11}. The channel behavior is captured through the parameter $p$, i.e., the probability of an ebit propagating through a quantum channel without experiencing absorption. And, $q \eqdef 1-p$ denotes the loss probability, i.e., the probability of ebit distribution failure as a consequence of the carrier absorption.

\begin{remark*}
    It is worthwhile to highlight that other noisy channel models can be easily incorporated in our analysis. As an example, Pauli channels followed by a purification process can be as well modeled with a Bernoulli distribution with parameter $p$, where $p$ denotes the success probability of the joint distribution and purification process.
\end{remark*}

We observe that, by exploiting heralded schemes, the super-node is able to recognize which client -- if any -- experienced an absorption over the channel. And, in case of absorption, further distributions can be attempted. Indeed, it may be necessary to attempt the distribution multiple times before having the targeted subset of clients in the network successfully received the ebit. From the above, it follows straightforward to consider, within our model, the \textit{number of possible distribution attempts} as the key temporal parameter. Clearly, the maximum number of distribution attempts is determined by the coherence times of the underlying quantum technology, as detailed in the next subsection. 

\subsection{Problem Formulation}
\label{sec:2.2}

\begin{figure*}[t!]
\centering
        \centering
        \includegraphics[width=0.9\textwidth]{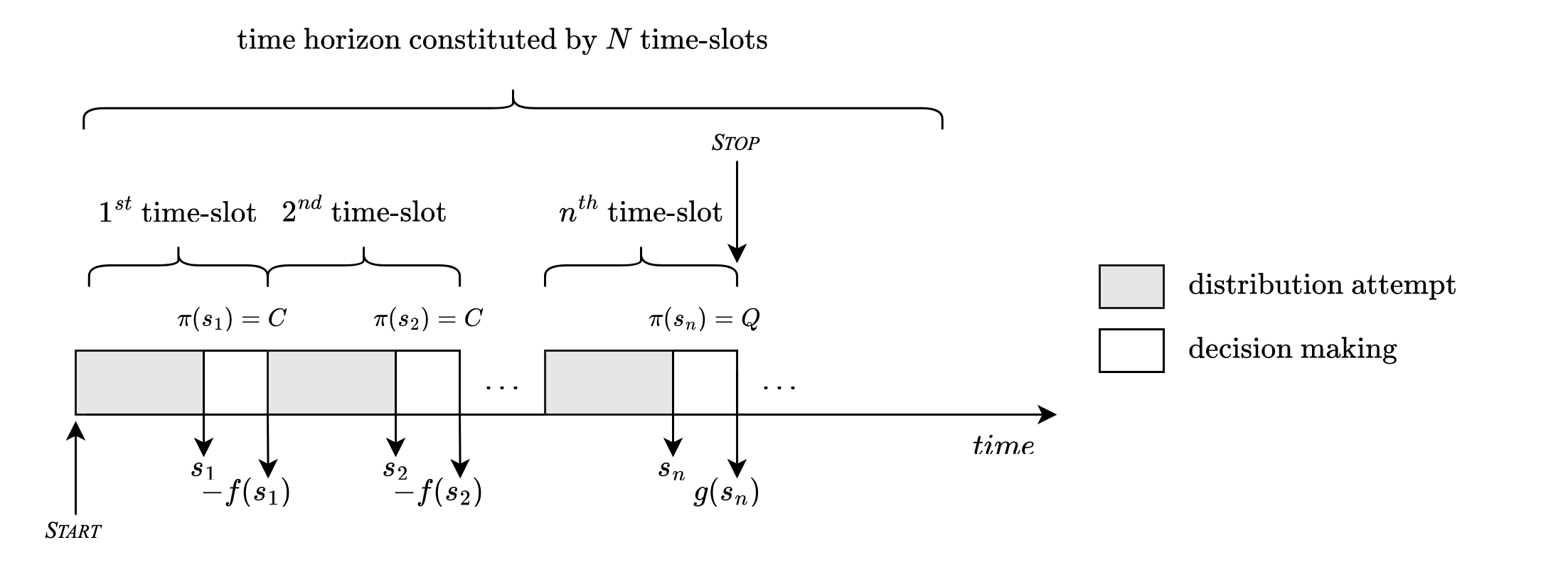}
        \caption{Pictorial representation of the model for the entanglement distribution process. The overall goal is to decide \textit{when to stop} the entanglement distribution.}
        \label{fig:01}
        \hrulefill
\end{figure*}

\begin{definition}[\textbf{Time horizon}]
\label{def:01}
    We consider the time horizon of the entanglement distribution process constituted by $N$ time slots:
    \begin{equation}
    \label{eq:01}
        \mathcal{N} = \{1,2, \dots, N \}.
    \end{equation}
    with $N$ implicitly accounting for the minimum guaranteed coherence time. 
\end{definition}

\begin{remark*}
    Specifically, the value of $N$ in \eqref{eq:01} depends on the particulars of the technology adopted for generating and distributing the entangled states, and it is set such that decoherence effects can be considered negligible within the time horizon.
\end{remark*}

\noindent As shown in Fig.~\ref{fig:01}, the time is organized into $N$ time-slots, where at (the end of) each time-slot the super-node can decide whether another distribution attempt should be performed (or not) in the subsequent time-slot. Clearly, the number of clients having already successfully received an ebit through the noisy channel, referred in the following as ``connected'' clients, represents a key parameter. We formalize this concept through the following two definitions.

\begin{definition}[\textbf{Action Set}]
\label{def:02}
    The \textit{action set} $\mathcal{A}$ denotes the set of actions available at the super-node:
    \begin{equation}
    \label{eq:02}
        \mathcal{A} = \{ C, Q \},
    \end{equation}
    with $C$ denoting the action of attempting another distribution round in the next time slot, and $Q$ denoting the action of not attempting the distribution. 
\end{definition}

\begin{definition}[\textbf{State Space}]
    \label{def:03}
    The system state space is defined as the pair 
     \begin{equation}
        \label{eq:03}
       (s,n) \in  \mathcal{\tilde{S}} \times \mathcal{N}, 
    \end{equation}
    where $\mathcal{N}$ is given in \eqref{eq:01} and $\tilde{S}$ is defined as follows:
    \begin{equation}
        \label{eq:04}
        \tilde{\mathcal{S}} \eqdef \mathcal{S} \cup \{\Delta\},
    \end{equation}
    with $\mathcal{S} \eqdef \{0,1,2, \dots, S \}$ denoting the set of possible values for number of connected clients. 
\end{definition}

\noindent Accordingly, the system is in state $(s,n)$ with $s \in \mathcal{S}$ if $s \leq S$ clients have successfully received an ebit from the super-node within the first $n$ distribution attempts. It is worthwhile to note that $\Delta$ in \eqref{eq:04} represents an auxiliary state, referred to as \textit{absorbing state}, that denotes the state of the system where no further distributions are attempted.  

\begin{remark*}
    In the following, we will use the symbols
    \begin{equation}
        \label{eq:05}
        s_n \eqdef (s,n)
    \end{equation} as a shorthand notation for the system state $(s,n)$, whenever this will not generate confusion. 
\end{remark*}
\begin{remark*}
     As mentioned, the overall goal is to distribute -- through teleportation -- a multipartite entangled state. However, the fidelity of a teleported quantum state increases with the fidelity of the distributed EPR pair, and deterministic quantum teleportation is achievable only by exploiting high fidelity EPR pairs. Remarkably, the proposed model allows to account for the fidelity of the distributed EPR pairs. More into details, whenever the fidelity $F$ of the distributed ebit results below a given fidelity threshold $F_{th}$, the distribution attempt is considered as failed. In fact, this event would prevent the correct teleportation of the multipartite entangled state. In this light, $p$ represents the probability of an ebit propagating through a quantum channel without experiencing absorption and received with fidelity above  the threshold $F>F_{th}$. Conversely, $q \eqdef 1-p$ denotes the probability of ebit distribution failure as a consequence of the carrier absorption, or as a consequence of the reception of an ebit with fidelity below the threshold $F<F_{th}$. Conversely, whenever an ebit is received with fidelity above the threshold $F>F_{th}$, the fidelity of the teleported multipartite state can be eventually further improved through \textit{entanglement purification} of the EPR pairs. Entanglement purification generally refers to the strategy able to obtain a single entangled state characterized by an higher fidelity from multiple imperfect entangled states\cite{BenBraPop-96}. Accordingly, entanglement purification demands for more than one ebit to be successfully distributed to each client. 
     Then, let $l$ be the number of ebits required at each node to perform entanglement purification and to distill one EPR with an higher fidelity value. 
     In such a case, our analysis continues to hold by considering an equivalent system composed by $\overline{S}=lS$ client nodes, each demanding for one successful ebit distribution.
\end{remark*}
\begin{remark*}
Entanglement purification implicitly assumes that each client node has at least $l$ communication qubits available at its side. Ideally, the super-node should distribute simultaneously $l$ ebits to each client. Clearly, this imposes additional requirements, as instance, on the number of communication qubits held at the super-node. Furthermore, the quantum channels should allow a distribution in batch, as instance with some sort of frequency-division strategy. Yet, whenever parallel distribution attempts are not allowed, the request of multiple ebit distribution per client imposes a delay in the distribution process as it demands for additional time-slots.         
\end{remark*}

\begin{definition}[\textbf{Allowed Action Set}]
\label{def:04}
    The allowed action set $A_{s_n}$ denotes the set of actions available at the super-node when the system state is $s_n$, and it results:
    \begin{equation}
        \label{eq:06}
        \mathcal{A}_{s_n} = 
        \begin{cases} 
            \{C,Q\} & s \in \mathcal{S}\setminus \{S\} \wedge  n<N \\
            \{Q\} & s = S \vee s = \Delta \vee n=N
        \end{cases}
    \end{equation}
\end{definition}
\noindent From Def.~\ref{def:04} it results that the only allowed action is $Q$ whenever the system either: i) successfully distributed entanglement to all the clients, or ii) is in the absorbing state $s=\Delta$, or iii) is at the last available time-slot $N$. Assuming the system being in the state $s_n \in \tilde{\mathcal{S}} \times \mathcal{N}$ and depending on the particular action $a \in \mathcal{A}_{s_n}$ taken, the system will evolve into some state $\tilde{s}_{n+1} \in \tilde{\mathcal{S}} \times \mathcal{N}$ with some probability $p(\tilde{s}_{n+1}|s_n,a)$, which will be derived with Lemma~\ref{lem:01} in Section~\ref{sec:3}. 

\begin{decfor}
During the first time-slot, the super-node simultaneously transmits $S$ ebits to the $S$ clients. In case of absorption, further distributions can be attempted. This requires additional time, thus challenging the decoherence constraints as well as impacting the overall distribution rate. Hence there exists a trade-off between the number of clients that successfully receive an ebit -- which we refer to as \textit{distributed cluster size} -- and the \textit{distribution time}, i.e., the number of time slots after which the distribution process is either completed or arrested. This trade-off deeply impacts the performance of the overlaying communication functionalities. Thus, its optimization becomes crucial in the design of quantum networks.
\end{decfor}

\noindent To capture this trade-off by abstracting from the particulars of the underlying hardware technology(ies), we model the effects of the action $a \in \mathcal{A}_{s_n}$, taken by the super-node  starting from the state $s_n$, through the notion of an utility function $r(s_n,a)$, referred to as \textit{reward function}. Accordingly, we formalize this concept in the following Definition.

\begin{definition}[\textbf{Reward function}]
    \label{def:05}
    Assuming that action  $a \in \mathcal{A}_{s_n}$ is taken when the system is in state $s_n \in \tilde{\mathcal{S}} \times \mathcal{N}$, the overall reward achieved is:
    \begin{equation}
        \label{eq:07}
        r(s_n,a) = \begin{cases}
            -f(s_n) & s \in \mathcal{S}, a = C \\
            g(s_n) & s \in \mathcal{S}, a= Q\\
            0 &  s= \Delta
        \end{cases}
    \end{equation}
    where:
    \begin{itemize} 
        \item[-] $f(s_n)$ denotes the \textit{continuation cost function}, which models the overall cost of attempting (continuing) the ebits distribution when the system is in $s_n$;
        \item[-] $g(s_n)$ denotes the \textit{pay-off function}, which models the gain achievable by stopping the ebits distribution when the system is in the state $s_n$.
    \end{itemize}
\end{definition}

\noindent It is clear that, according to our formulation, once the system reaches the absorption state, no further costs or rewards are obtained since the distribution process has been stopped.

\begin{remark*}
    The notion of reward function allows us to abstract from the particulars of i) the underlying technology for entanglement generation and distribution, and ii) the overlying network functionalities exploiting entanglement as a communication resource. In turn, this enables the following two key features: i) it restricts our attention on the effects of the entanglement distribution process; b) it allows us to measure the performance of an entanglement distribution strategy, and thus it allows us to quantitatively compare different strategies.
\end{remark*}

In the following we restrict our attention on payoff functions $\{g(s_n)\}$ satisfying the two following properties.

\begin{property}[\textbf{Monotonicity with $\textbf{s}$}]
    \label{prop:01}
The payoff function $g(s_n)$ is a monotonic non-decreasing function of $s$:  
    \begin{align}
        \label{eq:08}
         g(s_n) \leq g(\tilde{s}_n) \quad \text{with}\,\,\: s<\tilde{s}.
    \end{align}
\end{property}

\begin{property}[\textbf{Monotonicity with $\textbf{n}$}]
    \label{prop:02}
    The payoff function $g(s_n)$ is a monotonic non-increasing function of $n$:  
    \begin{align}
        \label{eq:09}
        g(s_n) \geq g(s_m) \quad \text{with}\,\,\:n \leq m.
    \end{align}
\end{property}

The rationale for these two properties is to model scenarios with meaningful meaning from an entanglement distribution perspective. Specifically, with Property~\ref{prop:01} the reward function tunes the system choice towards larger $s$, i.e., higher number of connected clients. Clearly, this is reasonable since the higher is the number of connected clients, the larger is -- as instance -- the distributed multipartite entangled state. Conversely, Property~\ref{prop:02} tunes the system choice towards shorter distribution times, which is mandatory to account for the fragile, easily degraded nature of entanglement.

\begin{remark*}
It is worthwhile to note that the theoretical framework developed in Sec.~\ref{sec:3.1} continues to hold regardless of whether the reward exhibits any monotonicity. Conversely, we will exploit these two properties in Sec.~\ref{sec:3.2} for reducing the computational complexity of the optimal decision strategy.
\end{remark*}

According to the theoretical framework developed so far, the entanglement distribution process is modeled through the quintuple:
\begin{equation}
    \label{eq:10}
    \{\tilde{\mathcal{S}}, \mathcal{N}, \mathcal{A}_{s_n}, p(\tilde{s}_{n+1}|s_n,a), r(s_n,a)\}. 
\end{equation}
\noindent The reader can refer to Table~\ref{tab:01} for a comprehensive summary of the notations used in the paper.

\section{Knowing when to Stop}
\label{sec:3}

Here, we develop the theoretical framework for modeling the entanglement distribution process. Specifically, in Sec.~\ref{sec:3.1}, we prove that -- with the minimal set of assumptions about the quantum technologies underlying entanglement generation and distribution -- the entanglement distribution process can be modeled as a Markov decision processes. Then in Sec.~\ref{sec:3.2} we prove some key properties that we will exploit to reduce the computational complexity of the problem.

\subsection{Optimal Decision Model}
\label{sec:3.1}
In Theorem~\ref{theo:01} we prove that the entanglement distribution process can be modeled as a Markov Decision Process. To this aim, the preliminary result in Lemma~\ref{lem:01} is needed.

\begin{lemma}
    \label{lem:01}
    Assuming action $a \in \mathcal{A}_{s_n}$ is taken when the system is in state $s_n \in \tilde{\mathcal{S}} \times \mathcal{N}$, the probability $p(\tilde{s}_{n+1}|s_n,a)$ of the system evolving into state $\tilde{s}_{n+1} \in \tilde{\mathcal{S}} \times \mathcal{N}$ depends only on current state and action, and it is given by:
    \begin{equation}
        \label{eq:11}
        p(\tilde{s}_{n+1}|s_n,a) = \begin{cases}
            p(\tilde{s}|s), &  \text{if} \,\, a = C \wedge s,\tilde{s} \in \mathcal{S} : \tilde{s} \geq s \\
            1 & \text{if} \,\, a = Q \wedge \tilde{s} = \Delta\\
            0 & \text{otherwise} 
        \end{cases},
    \end{equation}
    with
    \begin{equation}
        \label{eq:12}
        p(\tilde{s}|s)=\binom{S-s}{\tilde{s}-s}q^{S-\tilde{s}}p^{\tilde{s}-s}.
    \end{equation}
    \begin{IEEEproof}
        See Appendix~\ref{app:01}
    \end{IEEEproof}
\end{lemma}

\begin{figure*}
    \centering
    \hspace{-2 cm}
    \begin{subfigure}[b]{0.6\textwidth}
        \begin{tikzpicture}[->, >=stealth', auto, semithick, node distance=3cm
        ]
            \node[state]    (A) {$0$};
            \node[state]    (B) [right of=A] {$1$};
            \node[state]    (C) [right of=B] {$2$};
            \node[state]    (D) [right of=C] {$3$};
       	\draw
                (A) edge [loop below]   node{$q^3,C$} (A) 
                (A) edge [below right]   node[pos = 0.17]{$3p q^2,C$} (B)
                (A) edge [bend left]   node[pos = 0.54]{$3p^2 q,C$} (C)
                (A) edge [bend left = 42]   node{$p^3,C$} (D)
                (B) edge [loop below]   node{$q^2,C$} (B) 
                (B) edge [below right]   node[pos = 0.19]{$2pq,C$} (C)
                (B) edge [bend right = 30]   node[below]{$p^2,C$} (D)
                (C) edge [loop above]   node[above]{$q,C$} (C) 
                (C) edge [above right]   node[pos = 0.22]{$p,C$} (D);
        \end{tikzpicture}
    \subcaption{}{}
    \label{fig:02.a}
    \end{subfigure}
    \hspace{-0.4 cm}
    \begin{subfigure}[b]{0.3\textwidth}
        \begin{tikzpicture}[->, >=stealth', auto, semithick, node distance=2 cm,
        whitenode/.style={fill=white},
        greynode/.style={fill={rgb:black,1;white,7}}]
            \node[state,whitenode]    (A) at (0,-2) {$0$};
            \node[state,whitenode]    (B) at (1,0) {$1$};
            \node[state,whitenode]    (C) at (3,0)  {$2$};
            \node[state,whitenode]    (D) at (4,-2) {$3$};
            \node[state, greynode]    (E) at (2,-2){$\Delta$};
       	\draw
                (A) edge  node[sloped, below]{$1,Q$} (E)
                (B) edge  node[left]{$1,Q$} (E)
                (C) edge  node[right]{$1,Q$} (E)
                (D) edge  node[sloped, below]{$1,Q$} (E)
                (E) edge [loop below]  node{$1,Q$} (E);

            \matrix [below left] at (7.25,0.65) {
              \node [shape=circle,whitenode,draw=black,label=right:state in $\mathcal{S}$] {}; \\
              \node [shape=circle,greynode,,draw=black,label=right:absorbing state] {}; \\
            };
        \end{tikzpicture}
        \subcaption{}{}
        \label{fig:02.ab}
    \end{subfigure}
    \hfill
    \caption{
    Representation of the two functioning regimes for a network with $S=3$ clients: 
    (a): regime of the action $C$. 
    (b): regime of the action $Q$. 
    }
    \label{fig:02}
    \hrulefill
\end{figure*}
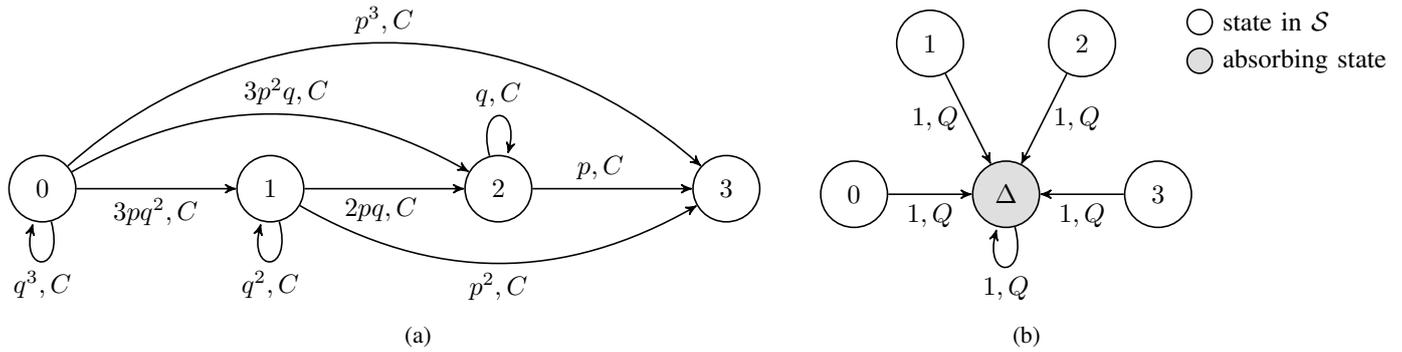

\begin{remark*}
    The available actions defined in \eqref{eq:06} establish two disjoint functioning regimes for the system, namely, the regime of action $C$ and the regime of action $Q$, as shown in Fig.~\ref{fig:02} with reference to a system with $S=3$ clients. Specifically, Fig.~\ref{fig:02.a} represents the regime of action $C$. Here, the system evolves according to the transition probabilities $p(\tilde{s}|s)$ in \eqref{eq:11}. It is worth noting that there exist no transition towards the absorbing state through action $C$. Differently, Fig.~\ref{fig:02.ab} represents the region of action $Q$. Specifically, by accounting for \eqref{eq:11}, once the super-node decides to perform action $Q$, the system will only evolve towards (or remain in) the absorbing state $\Delta$, where no further ebit transmissions are attempted. 
\end{remark*}

\begin{theorem}
    \label{theo:01}
    The entanglement distribution process can be modeled as a Markov Decision Process.
    \begin{IEEEproof}
        The proof follows from Lemma 1 by accounting for the Markov property of the transition probabilities \cite{Put-14}.
    \end{IEEEproof}
\end{theorem}

In the following, stemming from the result stated in Theorem~\ref{theo:01}, we will embrace the powerful framework of the Markov Decision Process to (optimal) ``\textit{know when to stop}'' the entanglement distribution process. To this aim, the following definition is needed.

\begin{definition}[\textbf{Policy}]
    \label{def:06}
    A policy $\pi(\cdot)$ is a rule determining the action to be taken in any possible state of the considered system. Hence, it is a function that maps the set of system states over the set of the allowed actions:
    \begin{equation}
        \label{eq:13}
            \forall s_n \in \mathcal{\tilde{S}} \times \mathcal{N}: \pi(s_n) \in \mathcal{A}_{s_n}
    \end{equation}
    In the following, $\Pi$ denotes the set of all possible policies.
\end{definition}

We note that, in \eqref{eq:13}, we exploited the Markovianity by considering policies $\pi(\cdot)$ depending on the current system state only, rather than on the entire history of the system state evolution \cite{Put-14}. Furthermore, we note that the overall reward achieved by adopting any policy $\pi(\cdot) \in \Pi$ is inherently stochastic, due to the noise affecting entanglement distribution. Thus, to assess and to compare the decision maker's preference toward different policies, we need a criterion to measure the performance of the selected policy. One widely adopted criterion in  literature is the \textit{expected total reward}, which we introduce in the following.

\begin{expectedRewards}
    Given that the strategy $\pi(\cdot)$ is adopted, the \textbf{total expected reward} $v_{\pi}(s_1)$, obtained when the system state starts in state $s_1$, is recursively defined as:
    \begin{equation}
        \label{eq:14}
            v_{\pi}(s_1) = r\big(s_1,\pi(s_1) \big) +
                \sum_{\tilde{s} \in \tilde{\mathcal{S}}: \tilde{s}_2=(\tilde{s}, 2) } p\big( \tilde{s}_2 | s_1 , \pi(s_1) \big)  v_{\pi}(\tilde{s}_2),
    \end{equation}
    \begin{figure*}
        \begin{equation}
            \label{eq:15}
                v_{\pi}(s_n) = \begin{cases}
                    \displaystyle r\big(s_n,\pi(s_n) \big) +
                        \sum_{\tilde{s} \in \tilde{\mathcal{S}}} p\big( \tilde{s}_{n+1} | s_{n} , \pi(s_{n}) \big)  v_{\pi}(\tilde{s}_{n+1}) & \text{ if } n<N \\
                        r\big(s_N, Q \big) & \text{otherwise}
                    \end{cases}
        \end{equation}
        \hrulefill
    \end{figure*}
    where $v_{\pi}(\tilde{s}_n)$ denotes the \textbf{expected remaining reward} at time slot $n$, and it is given by \eqref{eq:15} shown at the top of the next page. Specifically, the boundary condition at time slot $N$ in \eqref{eq:15} prevents from infinite loops in the absorbing state.
\end{expectedRewards}

We note that, for deriving the expression in \eqref{eq:14}, we exploited Theorem~4.2.1 in \cite{Put-14}. Accordingly, it is possible to restrict our attention on deterministic policies $\pi(\cdot) \in \Pi$ with no loss of optimality. Furthermore, we note that the expected total reward $v_{\pi}(s_1)$ has been defined as a recursive function, where the recursive step $v_{\pi}(s_n)$ at time slot $n$ is function of three key parameters. That are the number of connected clients $s$, the policy $\pi(\cdot)$ through action $\pi(s_{n})$, and the reward at time slot $n+1$ via the transition probabilities $p\big( \cdot| s_n, \pi(s_{n}) \big)$.

Stemming from the above, we are ready now to formally define the problem of (optimal) \textit{knowing when to stop} the entanglement distribution.

\begin{whenToStop}
    By accounting for \eqref{eq:14}, the overall objective is to find the strategy $\pi^* \in \Pi$ that maximizes the expected total reward when the system is in state $s_1$:
    \begin{equation}
        \label{eq:16}
       v_{\pi^*}(s_1) = \max_{\pi \in \Pi} \big\{ v_{\pi}(s_1) \big\}
    \end{equation}
\end{whenToStop}

\noindent As a matter of fact, being the considered sets $\tilde{\mathcal{S}}$ and $\mathcal{N}$ finite, there always exists a deterministic strategy achieving the maximum in \eqref{eq:16} \cite{Put-14}. Furthermore, we have implicitly assumed as overall goal to maximize the reward for some specific initial state $s_1$. Alternatively, the goal might be to find the optimal policy $\pi^*$ prior to know the initial state $s_1$. In such a case, by accounting for \eqref{eq:14}, the total expected reward $v_{\pi}$ is given by:
\begin{equation}
    \label{eq:17}
    v_{\pi} = \sum_{s \in \mathcal{S}: s_1=(s, 1) } p(s) v_{\pi}(s_1)
\end{equation}
with $p(s)$, namely, the probability of successfully distributing ebits to $s$ clients during the first distribution attempt, given by:
\begin{equation}
    \label{eq:18}
    p(s)=p^s q^{S-s}
\end{equation}
However, the reward in \eqref{eq:17} is maximized by maximizing the reward in \eqref{eq:14} for each $s_1$ in $\mathcal{S}$ \cite{Put-14}. Hence in the following we will focus on the problem formulation in \eqref{eq:16} without any loss in generality.

\subsection{Optimal Decision Strategy: Properties}
\label{sec:3.2}

In this subsection, we prove that the optimal policy $\pi^*(\cdot)$ exhibits specific properties with respect to the reward function. Then, we will engineer these properties to derive effective, practical strategies for reducing the computational complexity of the decision problem. To this aim, some preliminaries are needed.

First, we explicit the expression of the expected remaining reward in \eqref{eq:15}. Specifically, let us denote with $v^*(s_1)$ the maximum expected total reward, which is equivalent to the expected total reward achieved by the optimal policy $\pi^*$ given in \eqref{eq:16}:
\begin{equation}
    \label{eq:19}
    v^*(s_1) \eqdef v_{\pi^*}(s_1) 
\end{equation}
By accounting for the allowed action set $\mathcal{A}_{s_n}$ given in \eqref{eq:06} and for the reward function defined in Def.~\ref{def:05}, the maximum expected total reward $v^*(s_1)$ is given in \eqref{eq:20} shown at the top of the next page, with the maximum expected remaining reward at the $n$-th recursive step given by:
\begin{figure*}
    \begin{align}
        \label{eq:20}
        v^*(s_1) & \displaystyle = \max \left\{
            \overbrace{r(s_1,Q)}^{\eqdef v^*_Q(s_1)} \, , \,
            \overbrace{r\big(s_1, C \big) +
            \sum_{\tilde{s} \in \tilde{\mathcal{S}}} p\big( \tilde{s}_{2} | s_1 , C) \big)  v^*(\tilde{s}_2)}^{\eqdef v^*_C(s_1)}
        \right\} = \max \displaystyle\left\{
            g(s_1) \, , \,
            -f(s_1) +
            \sum_{\tilde{s} \in \tilde{\mathcal{S}}} p\big( \tilde{s}_{2} | s_1 , C) \big)  v^*(\tilde{s}_2)
        \right\}
    \end{align}
    \hrulefill
\end{figure*}
\begin{equation}
    \label{eq:21}
    v^*(s_n) = \begin{cases}
            \max \left\{  v^*_Q(s_n), v^*_C(s_n) \right\} & \text{if } n < N \\
            r(s_N,Q) & \text{otherwise}
        \end{cases}
\end{equation}
In \eqref{eq:20}, $v^*_Q(s_1)$ and $v^*_C(s_1)$ denote the maximum expected reward achievable when action $Q$ or $C$ is taken, respectively, starting from state $s_1$.

Furthermore, let us denote with $p(\breve{s}_{n+k}|s_n,C)$ the probability to evolve into state $\breve{s}_{n+k}= (\breve{s}, n+k)$ at time slot $n+k$, starting from state $s_n=(s,n)$ with $s \neq \Delta$, by having chosen always action $C$ at the end of each of time-slot\footnote{Namely, by choosing action $C$ regardless whether the number of connected clients $s$ is either $s < S$ or $s = S$.} between $n$ and $n+k-1$. By exploiting the Markovianity in Lemma~\ref{lem:01}, this probability, referred to us \textit{extended transition probability}, can be recursively written as follows:
\begin{equation}
    \label{eq:22}
    p(\breve{s}_{n+k}|s_n,C)= \sum_{\tilde{s}=s}^{\breve{s}} p\left( \breve{s}_{n+k}|\tilde{s}_{n+1},C\right) p\left(\tilde{s}_{n+1} |s_n, C \right),
\end{equation}
with the expression of $p\left(\tilde{s}_{n+1} |s_n, C \right)$ given in Lemma~\ref{lem:01}.

Stemming from the extended transition probabilities given in \eqref{eq:22}, we are ready to define now two rewards functions, that will be exploited in the following for efficiently deriving the optimal policy.
\begin{rewardMM}
    \label{rewardMM}
    Given that the system is in state $s_n=(s,n)$, with $s\neq \Delta$ and $n<N$, we introduce the quantities $v^+(s_n)$ and $v^{-}(s_n)$, referred to as the reward majorant and the reward minorant, respectively:
    \begin{align}
        \label{eq:23}
        v^+(s_n) &= r\big(s_n,C \big) +
                        \sum_{\breve{s} \in \tilde{\mathcal{S}}} p\big( \breve{s}_{N} | s_{n} , C \big) v_Q^*\big(\breve{s}_{n+1} \big)\\&\nonumber
                        = -f(s_n) +
                        \sum_{\breve{s} \in \tilde{\mathcal{S}}} p\big( \breve{s}_{N} | s_{n} , C \big) g(\breve{s}_{n+1})\\
        \label{eq:24}
        v^-(s_n) &= r\big(s_n,C \big) +
                        \sum_{\tilde{s} \in \tilde{\mathcal{S}}} p\big( \tilde{s}_{n+1} | s_{n} , C \big) v_Q^*\big(\tilde{s}_{n+1} \big)=\\&\nonumber
                        = -f(s_n) +
                        \sum_{\tilde{s} \in \tilde{\mathcal{S}}} p\big( \tilde{s}_{n+1} | s_{n} , C \big) g(\tilde{s}_{n+1}),
    \end{align}
\end{rewardMM}
with $\tilde{s}_{n+1} = (\tilde{s}, n+1)$ and $\breve{s}_{n+1}  = (\breve{s}, n+1)$.

Both the majorant and the minorant model the reward achievable by deciding first to continue the entanglement distribution at time slot $n$ and, then, to stop the distribution at the subsequent time slot $n+1$. Yet, they significantly differ each other:
\begin{itemize}
    \item[-] The reward minorant $v^-(s_n)$ is obtained by assuming the system evolving from state $s_n$ to state $\tilde{s}_{n+1}$ in agreement with the transition probabilities given in \eqref{eq:13}.
    \item[-] Conversely, the reward majorant $v^+(s_n)$ is obtained by assuming the system able to evolve freely from state $s_n$ to state $\breve{s}_{N}$ -- with $\breve{s}_{N}=(\breve{s},N)$ representing the state that would have been reached by performing $N-n$ subsequent distributions attempts by choosing only action $C$ and never action $Q$ -- yet in a single time slot. In other words, the majorant models the expected reward achieved when the system performs $N-n$ subsequent distributions attempts, yet i) by paying only a single continuation cost $-f(s_n)$, and ii) by obtaining a pay-off $g(\breve{s}_{n+1})$ as if $\breve{s}$ would have been reached in a single time slot.
\end{itemize}

The proof of the main result, namely, Theorem~\ref{theo:02} requires the following preliminary lemma.

\begin{lemma}
    \label{lem:02}
    Given that the system state is $s_n$ with $s \in \mathcal{S}$ and $n < N$, it results:
    \begin{equation}
            \label{eq:25}
            v^-(s_n) \leq v^*_C(s_n) \leq v^+(s_n)
    \end{equation}
    \begin{IEEEproof}
        See Appendix~\ref{app:01}.
    \end{IEEEproof}
\end{lemma}

\begin{theorem}
    \label{theo:02}
    Given that the system state is $s_n$ with $s \in \mathcal{S}$ and $n < N$, it results:
    \begin{equation}
            \label{eq:26}
            \pi^*(s_n) = \begin{cases}
                    Q & \text{if } g(s_n) \geq v^+(s_n) \\
                    C & \text{if } g(s_n) \leq v^-(s_n)
                \end{cases}
    \end{equation}
    \begin{IEEEproof}
        The proof follows directly from Lemma~\ref{lem:02}, by accounting for the definition of $v^*_C(s_n)$ and $v^*_Q(s_n)$ given in \eqref{eq:20}.
    \end{IEEEproof}        
\end{theorem}

Markov decision problems as the one we considered in \eqref{eq:16} are generally solved with backward induction \cite{Put-14}. Specifically, stemming from the expression of the maximum expected remaining reward given in \eqref{eq:21}, backward induction works as follows: starting from $n=N$ and going backward in time, the optimal action maximizing the expected total reward is obtained for each state $s_n$ by exploiting the already-derived optimal actions for states $\tilde{s}_{n+1}$, with $\tilde{s} > s$.

\begin{remark*}
    When the system state is $s_n$, backward induction requires to preliminarly evaluate $\big( S-s+1\big)^{N-n}$ optimal actions -- i.e., to compute the optimal action for each possible future state -- before determining the optimal action $\pi^*(s_n)$ for the current state. Luckily, with Theorem~\ref{theo:02} we have derived an efficient strategy for finding the optimal action without the need of evaluating the future evolution of the system. Specifically, whenever $g(s_n)$ satisfies one of the conditions in \eqref{eq:26}, the optimal action can be decided regardless of any further evolution of the system. We validate this result with the first experiment in Sec.~\ref{sec:4}.
\end{remark*}

Finally, it is important to discuss the assumptions underlying Theorem~\ref{theo:02}. As regards to the continuation cost $f(\cdot)$, Theorem~\ref{theo:02} does not require any assumption or constraint, except $f(\cdot)$ being reasonable non negative\footnote{Otherwise it would represent a pay-off rather than a cost.}. As regards to the pay-off function $g(\cdot)$, Theorem~\ref{theo:02} requires Properties~\ref{prop:01}-\ref{prop:02} being satisfied. Yet these properties are not restrictive, since they reasonably drive the entanglement distribution toward entangling the \textit{larger} number of client nodes in the \textit{shorter} possible time-frame.

In the next subsection, we will introduce and discuss some (reasonable) assumptions on the pay-off function which allows us to further simplify the search of the optimal policy.

\subsection{One-Step Look Ahead}
\label{sec:3.3}

Here we depart from the general discussion of Sec.~\ref{sec:3.2}, by further extending the result of Theorem~\ref{theo:02} for deriving the optimal policy, albeit imposing additional constraints on the rewards. To this aim, the following preliminaries are need.

Given that there exists only two actions in \eqref{eq:02} -- namely, continue or stop -- the entanglement distribution problem belongs to the framework of optimal \textit{stopping} problems, for which there exists a very simple (hence, computational efficient) rule -- namely, one-step look ahead (OLA) rule --  for deciding the action to be taken.

\begin{definition}[\textbf{OLA Set}]
	\label{def:07}
	At time-step $n$, the one-step look ahead (OLA) set $\mathcal{S}^{Q}_n \subseteq \tilde{\mathcal{S}}$ is the set of system states where the instantaneous reward achievable by stopping is not lower than the expected reward achievable by attempting a further distribution attempt and then deciding to stop the distribution.
	\begin{equation}
		\label{eq:27}
		\mathcal{S}^{Q}_n = \left\{ s \in \mathcal{S} : g(s_n) \geq v^-(s_n) \right\} 
	\end{equation}
	with $v^-(s_n)$ given in \eqref{eq:24}.
\end{definition}

\begin{definition}[\textbf{OLA Rule}]
	\label{def:08}
	\begin{equation}
		\label{eq:28}
		\pi(s_n) = \begin{cases}
				Q & \text{if } s_n \in \mathcal{S}^{Q}_n \Longleftrightarrow g(s_n) \geq v^-(s_n) \\
				C & \text{otherwise}
			\end{cases}
	\end{equation}
\end{definition}

The naming for the OLA rule follows by noting that the reward minorant $v^-(s_n)$ represents the \textit{expected reward} when the policy is \textit{to continue for one-step and then to stop}, namely:
\begin{equation}
	\label{eq:29}
	v^-(s_n) = -f(s_n) + E[g(\mathfrak{S}_{n+1})]
\end{equation}
with $\mathfrak{S}_{n+1}$ denoting the random variable describing the system state at step $n+1$.

The OLA rule is optimal whenever the OLA set is closed \cite{Hameed-77, Yasuda-88}, namely, whenever the system state remains confined within the OLA set, once entering. Unfortunately, the optimality of the OLA rule strictly depends on the particulars of the cost $f(\cdot)$ and pay-off $g(\cdot)$ functions, and no general conclusions can be taken independently.

Yet, we can consider different settings for the cost/pay-off functions -- which allows us to model a wide range of possible communication scenarios -- and discuss the optimality of the OLA rule with respect to this setting. More into details, we consider the following three base-cases:
\begin{align}
	\label{eq:30}
	g(s_n) &= \frac{s}{n} \\
	\label{eq:31}
	g(s_n) &= \lambda^n s, \text{ with } \lambda \in (0,1] \\
    \label{eq:32}
	g(s_n) &= \frac{s}{S} - \frac{n}{N}
\end{align}
with $f(s_n) = 0$ since we already incorporated the cost arising with additional distribution attempts into the reward.

\begin{remark*}
    As an example, with the first base-case given in \eqref{eq:30} we model a scenario where the reward, represented by the number $s$ of entangled clients, is discounted by a factor equal to the number of time-slots used for entangling such clients. The rationale for this scenario is to model the reward as a sort of \textit{entanglement throughput} -- namely, as an    \textit{average entanglement per unit of time} -- similarly to the bit throughput that represents one of the key metric for classical networks. As regards to the second base-case given in \eqref{eq:31}, it introduces a discount factor $\lambda$ which exponentially weights the reward $s$ as time passes. As a matter of fact, multiplicative decreasing the rate of some process such as in \eqref{eq:31} is widely adopted in classical networks, with TCP exponential back-off constituting the most famous case. Finally, with \eqref{eq:32} we meant to introduce another base-case for conferring generality to the discussion.
\end{remark*}

By considering the settings of the base-cases, we have the following result.

\begin{prop}
    When the rewards are modeled as in \eqref{eq:31} or \eqref{eq:31}, the OLA rule is optimal and it results:
    \begin{equation}
        \label{eq:33}
        \pi^*(s_N) = Q \Longleftrightarrow \begin{cases}
                \displaystyle s \geq \frac{\lambda Sp}{1- \lambda + \lambda p} \displaystyle & \text{if } g(s_n) = \lambda^n s \\
                s \geq S - \frac{S}{Np} & \text{if } g(s_n) = \frac{s}{S} - \frac{n}{N}
	       \end{cases}
    \end{equation}
    whereas when the rewards are modeled as in \eqref{eq:30}, the OLA rule is not optimal.
    \begin{IEEEproof}
        See Appendix~\ref{app:03}.
    \end{IEEEproof}
\end{prop}

From an engineering perspective, it is evident that having an efficient (i.e., low-computational-complexity) optimal rule, such as the OLA rule, for deriving the optimal policy -- namely, for deciding when to stop distributing entanglement within a quantum network -- is highly advantageous. Hence, whenever possible, the opportunity of choosing rewards satisfying the optimality condition of the OLA rule should be preferred.

Nevertheless, whenever this should not be possible, we can still exploit the main result -- namely, Theorem~\ref{theo:02} -- for designing an efficient rule, as long as we tolerate finding a sub-optimal policy rather than an optimal one.

\begin{definition}[\textbf{Sub-Optimal Rule}]
	\label{def:09}
	\begin{equation}
		\label{eq:34}
		\pi(s_n) = \begin{cases}
				\displaystyle Q & \text{if } s_n \geq \frac{v^+(s_n) + v^-(s_n)}{2} \\
				C & \text{otherwise}
			\end{cases}
	\end{equation}
\end{definition}

Clearly, the ``amount'' of sub-optimality -- hence, the loss in reward -- introduced by such a rule strictly depends on the particular settings of the rewards. In the next subsection, we will evaluate such a sub-optimality for the three base-cases introduced above.

\section{Performance evaluation}
\label{sec:4}

In this section, we first validate the theoretical results derived in Secs.~\ref{sec:3.2} and \ref{sec:3.3}.

Then, we discuss the impact of the reward functions on the performance of the entanglement distribution process. To this aim, we focus on two key metrics:
\begin{itemize}
    \item average distribution time, namely, the average number of time-slots before the distribution is arrested;
    \item average cluster size, namely, the average number of client nodes successfully entangled;
\end{itemize}
More into details, we investigate how the choice of the reward setting influences these two key metrics. This allows us to to draft some guidelines for selecting a reward function able to drive the system to fulfill some specific performance requirements.

\begin{figure*}[t]
    \begin{minipage}[t] {0.49\textwidth}
        \centering
        \includegraphics[width=\columnwidth]{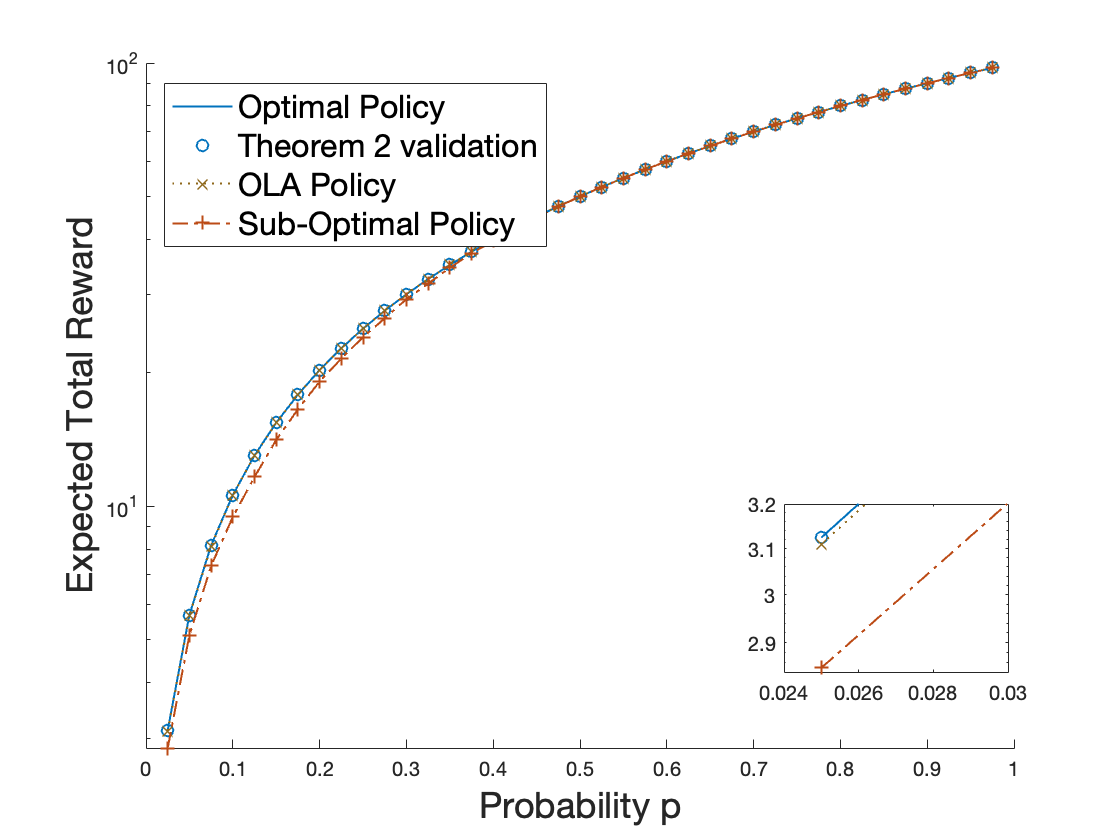}
	    \caption{Expected total reward $v_{\pi}$ as a function of the ebit propagation probability $p$ for $S = 100$, $N=100$ and $g(s_n) = \frac{s}{n}$. Logarithmic scale for axis $y$.}
	    \label{fig:03}
    \end{minipage}
    \hspace{0.02\textwidth}
    \begin{minipage}[t]{0.49\textwidth}
        \centering        \includegraphics[width=\columnwidth]{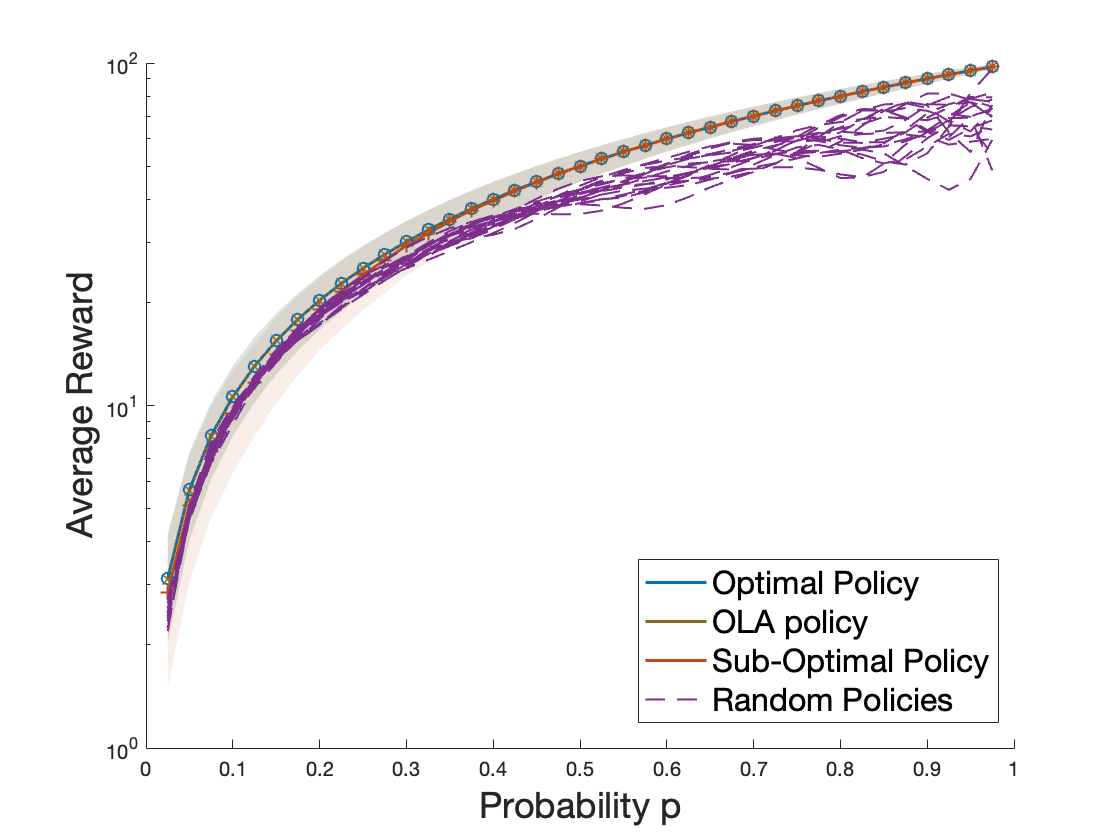}
	    \caption{Average total reward $v_{\pi}$ as a function of the ebit propagation probability $p$ for the same setting of Fig.~\ref{fig:03}. Lines denote the mean, whereas shading areas denote the standard deviation. Logarithmic scale for axis $y$.}

	\label{fig:04}
    \end{minipage}
    \hrulefill
\end{figure*}

With the first experiment, we evaluate in Fig.~\ref{fig:03} the expected total reward $v_{\pi}$ given in \eqref{eq:17} as a function of the ebit propagation probability $p$. The adopted simulation set is as follows: the number of clients is $S=100$, the time-horizon is constituted by $N=100$ time-slots, the rewards are modeled as in \eqref{eq:30} with $g(s_n) = \frac{s}{n}$, and $p$ varies with step equal to $0.025$. Within the experiment, we consider four different rewards.

\noindent First, we consider the reward $v_{\pi^*}$ achieved with the optimal policy $\pi^*$, with $\pi^*$ obtained via exhaustive search through backward induction. Clearly, this is the maximum expected reward that can be achieved, and it represents the performance baseline for any sub-optimal policy. We note that, the higher is $p$, the higher is the reward $v_{\pi^*}$. This result is reasonable, since higher distribution probabilities allow the system to evolve toward states characterized by higher cluster sizes $s$ and lower distribution times $n$.

\noindent Additionally, we consider the reward $v_{\pi^*}$ achieved with the policy $\pi^*$ computed via Theorem~\ref{theo:02}. More into detail, $\pi^*(s_n)$ is obtained with Theorem~\ref{theo:02} whenever either of the two constrains in \eqref{eq:26} holds, and via backward induction otherwise. Clearly, by comparing this reward with the optimal reward $v_{\pi^*}$, we can observe a perfect agreement between the two rewards. This constitutes an experimental validation of the analytical results derived in Theorem~\ref{theo:02}.

\noindent Furthermore, we consider the reward $v_{\pi}$ achieved when the policy $\pi$ is obtained with the OLA rule given in Definition~\ref{def:08}. Indeed, it must be noted that -- although barely noticeable even in the zoomed-in inset of Fig.~\ref{fig:03} -- the reward achievable with the OLA rule is lower than the reward $v_{\pi^*}$ achievable with the optimal policy for any value of $p$. This validates the theoretical results derived in Prop~\ref{prop:01}, and, specifically, the sub-optimality of the OLA rule for $g(s_n) = \frac{s}{n}$. Yet, the performance degradation of the OLA rule is practically negligible. 

\noindent Finally, we consider the reward $v_{\pi}$ achieved with the policy $\pi$ obtained via the sub-optimal rule given in Definition~\ref{def:09}. From Fig.~\ref{fig:03}, one might question the rationale for this sub-optimal rule and, specifically, one might incorrectly believes that -- given that the OLA rule significantly outperforms the sub-optimal rule given in Definition~\ref{def:09} -- the last rule is useless. Yet it must be noted that the performance of the OLA rule strictly depends on some specific assumptions on the cost $f(\cdot)$ and pay-off $g(\cdot)$ functions, assumptions which are not required by the rule given in Definition~\ref{def:09}.

\begin{remark*}
From the above discussion, it becomes clear that there exists a trade-off between optimality and computational-efficiency, that must be properly engineered by the quantum network designers. Specifically, designers can decide to adopt generalist heuristic policies -- such as the one in Def.~\ref{def:09}  -- 
which does not impose limitations on the choice of the reward functions albeit at the price of sub-optimal decisions. Or they can leverage optimal, efficient policies -- such as the OLA one -- as long as they can tolerate additional constraints in the reward function definition.
\end{remark*}

With the second experiment, we aim at assessing the importance of an optimal policy for achieving the highest total reward. For this, in Fig.~\ref{fig:04} we plot the average total reward $v_{\pi}$ given in \eqref{eq:17} as a function of the ebit propagation probability $p$ for $10^6$ Montecarlo distribution process trials, for the same simulation set adopted in Fig.~\ref{fig:03}. We note that lines denote the mean of the total rewards over the different trials, whereas shading areas denote the standard deviation of the different trials\footnote{With the shading areas of optimal and OLA rewards practically overlapping.}.

\noindent We extend the set of policies by considering -- along with the optimal and the two sub-optimal policies already considered in the previous experiment -- 20 random policies. We observe that the higher is the ebit distribution probability $p$ , the higher is the performance gap between the expected total reward achieved by the optimal strategy and the reward achieved by a random strategy. As a matter of fact, the performance gap remains evident even if we consider the distribution of the optimal reward via standard deviation. This result shows the importance of the considered problem for scenarios of practical interest, namely, for scenarios where entanglement can be fairly distributed. 

\begin{figure*}[t]
    \begin{minipage}[t] {0.49\textwidth}
        \centering
        \includegraphics[width=\columnwidth]{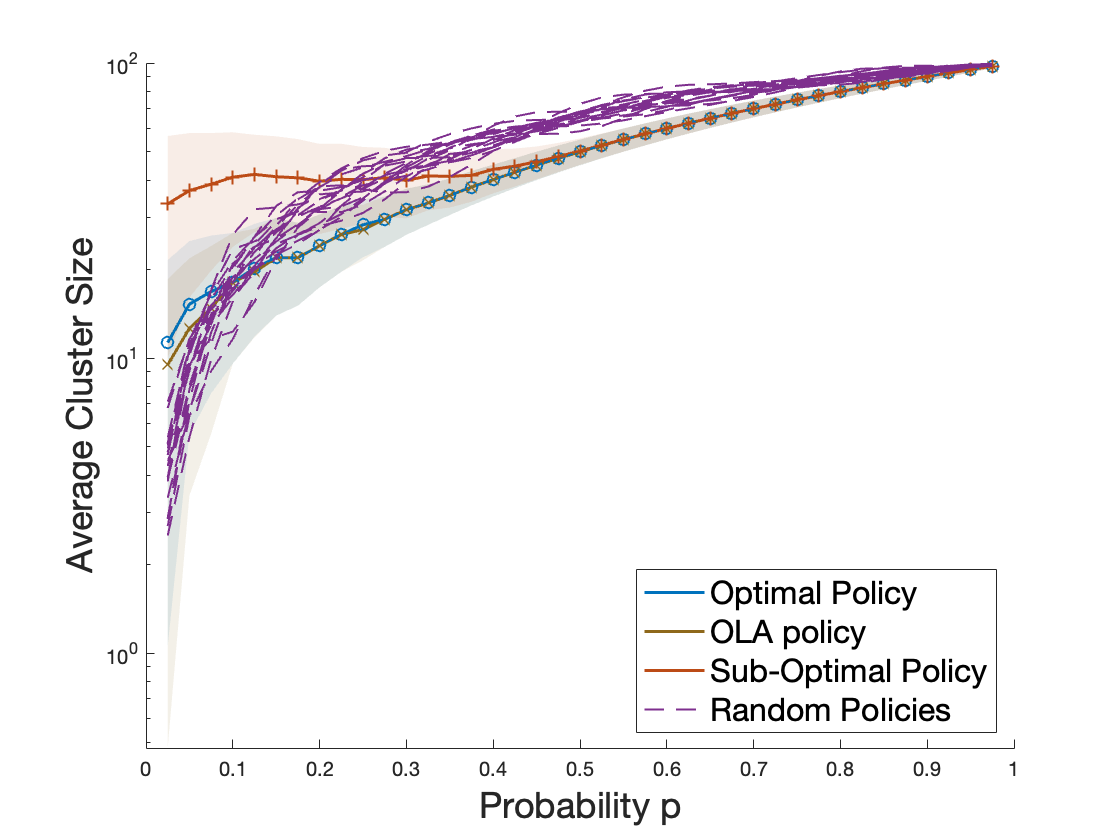}
	    \caption{Average cluster size $s$ as a function of the ebit propagation probability $p$ for the same setting of Fig.~\ref{fig:03}. Lines denote the mean, whereas shading areas denote the standard deviation. Logarithmic scale for axis $y$.}
	    \label{fig:05}
    \end{minipage}
    \hspace{0.02\textwidth}
    \begin{minipage}[t]{0.49\textwidth}
        \centering
        \includegraphics[width=\columnwidth]{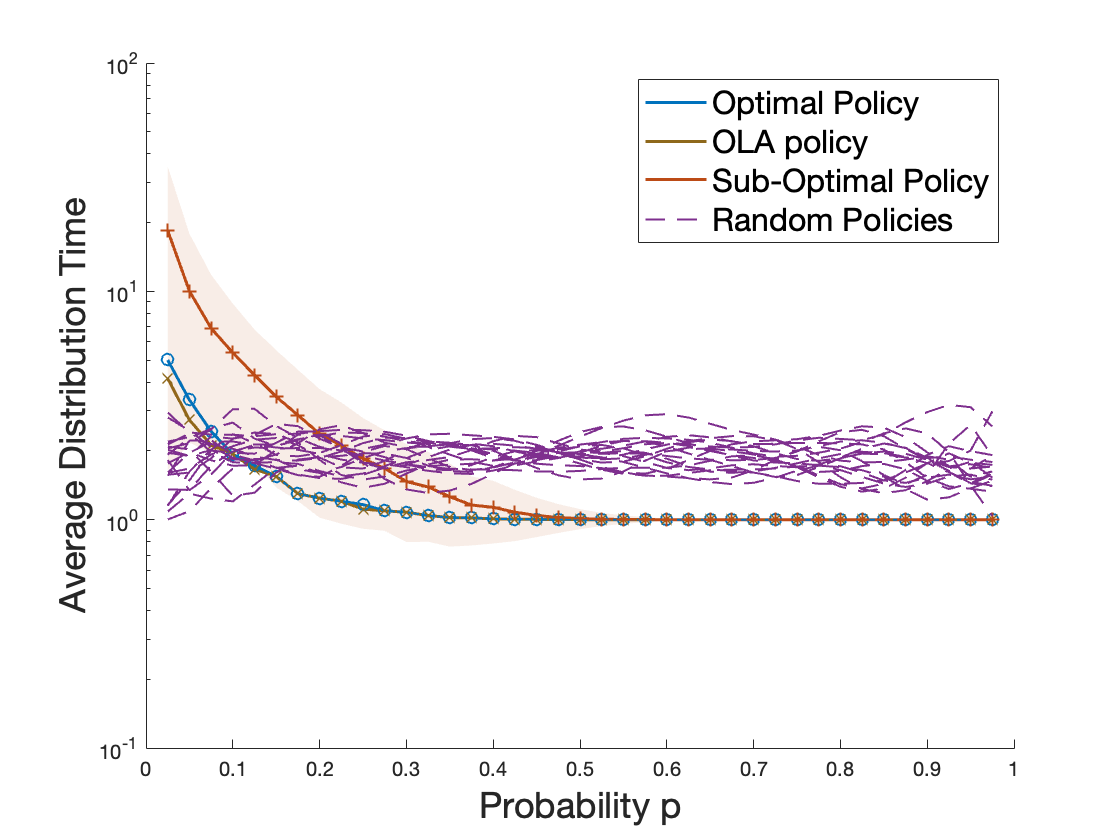}
	    \caption{Average distribution time $n$ as a function of the ebit propagation probability $p$ for the same setting of Fig.~\ref{fig:03}. Lines denote the mean, whereas shading areas denote the standard deviation. Logarithmic scale for axis $y$.}

	\label{fig:06}
    \end{minipage}
    \hrulefill
\end{figure*}

In Fig.~\ref{fig:05}, we present the average cluster size $s$ as a function of the ebit propagation probability $p$, computed with the same $10^6$ Montecarlo distribution process trials of Fig.~\ref{fig:04}. As before, lines denote the mean over the different trials, whereas shading areas denote the standard deviation of the different trials. First, we note that the random policies might achieve larger cluster sets with respect to the optimal policy. The rationale for this behaviour is that the optimal policy aims at: i) maximizing the cardinality of the cluster set, while simultaneously ii) minimizing the distribution time. Hence, depending on $g(\cdot)$ and $p$, the optimal policy might prefer an earlier stop of the distribution process. And this was, indeed, the overall objective of our modeling.

These considerations are confirmed by Fig.~\ref{fig:06}, which presents the average distribution time $n$ as a function of ebit propagation probability $p$, computed with the same $10^6$ Montecarlo distribution process trials of Fig.~\ref{fig:04}. Indeed, it is possible to note that the values of $p$ in Fig.~\ref{fig:06} -- for which the random policies achieve larger cluster sizes with respect to the optimal policy -- are characterized by longer distribution times.

\begin{figure*}[t]
    \begin{subfigure}[t] {0.32\textwidth}
        \centering
        \includegraphics[width=\columnwidth]{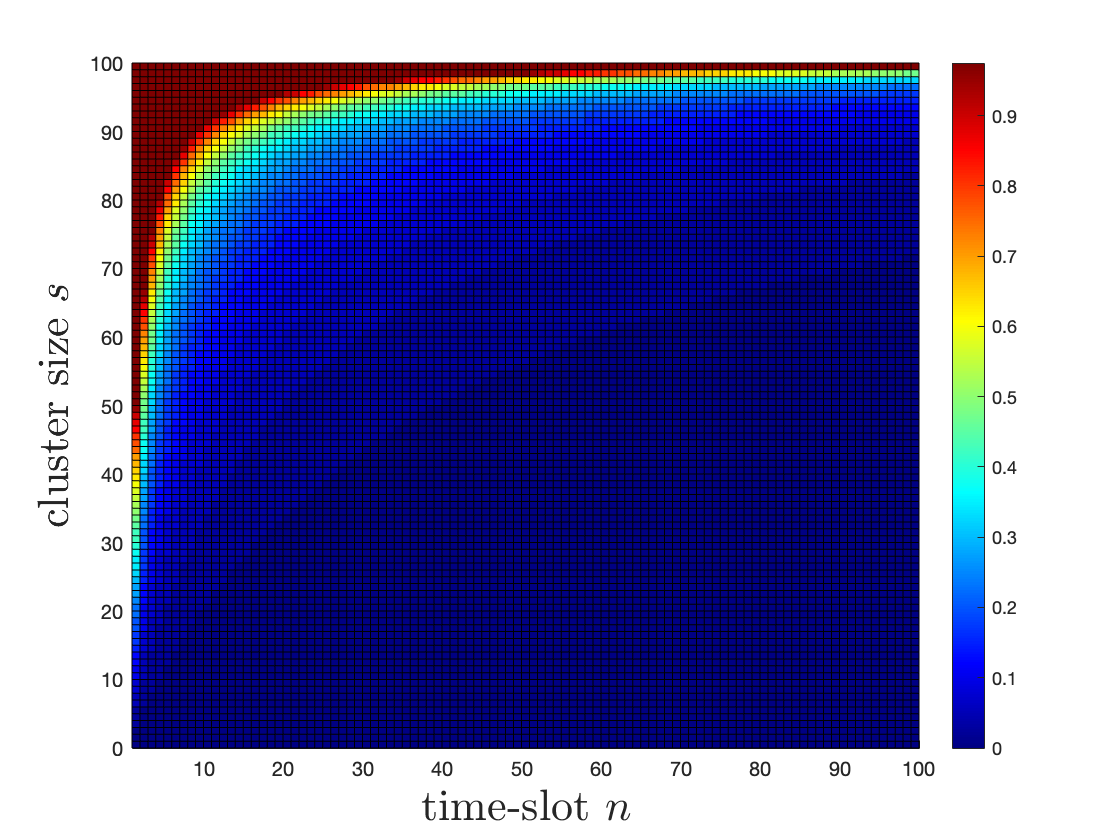}
	    \subcaption{Action matrix for pay-off function $g(s_n) = \frac{s}{n}$.}
	    \label{fig:09-a}
    \end{subfigure}
    \hspace{0.02\textwidth}
    \begin{subfigure}[t]{0.32\textwidth}
        \centering
        \includegraphics[width=\columnwidth]{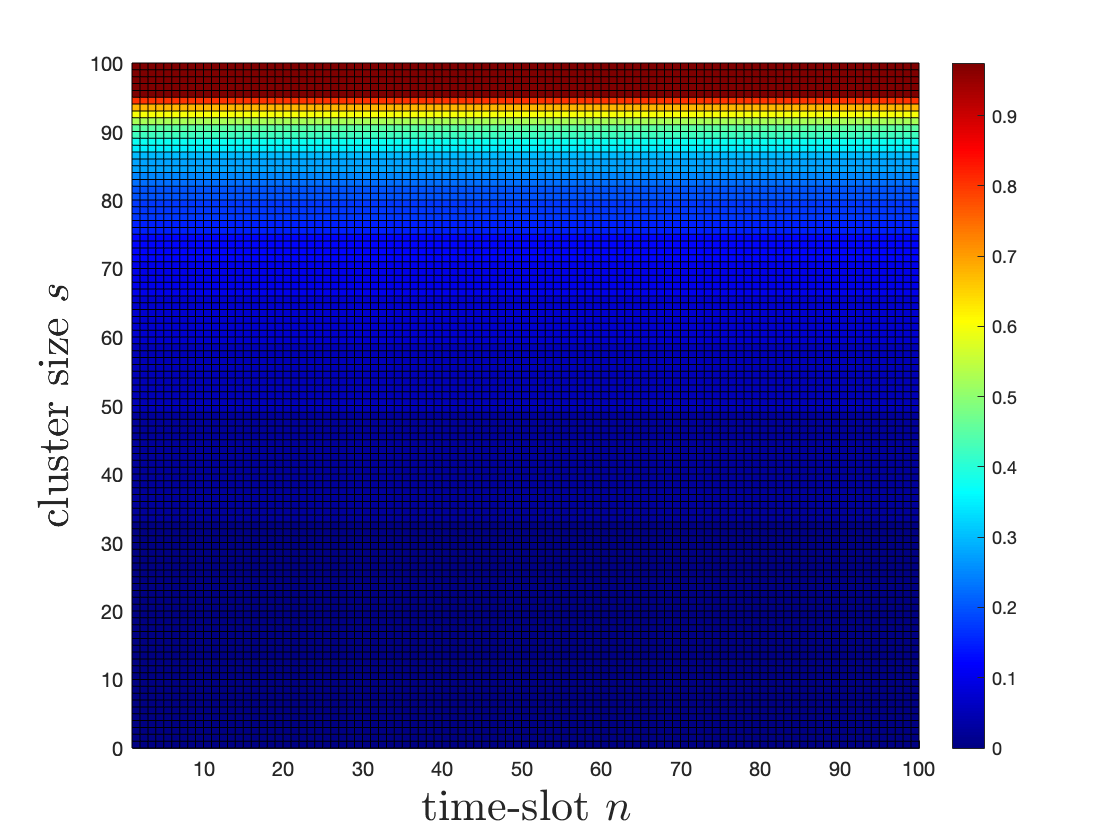}
	    \subcaption{Action matrix for pay-off function $g(s_n) = \lambda^n s$, with $\lambda = 0.95$.}
        \label{fig:09-b}
    \end{subfigure}
    \hspace{0.02\textwidth}
    \begin{subfigure}[t]{0.32\textwidth}
        \centering
        \includegraphics[width=\columnwidth]{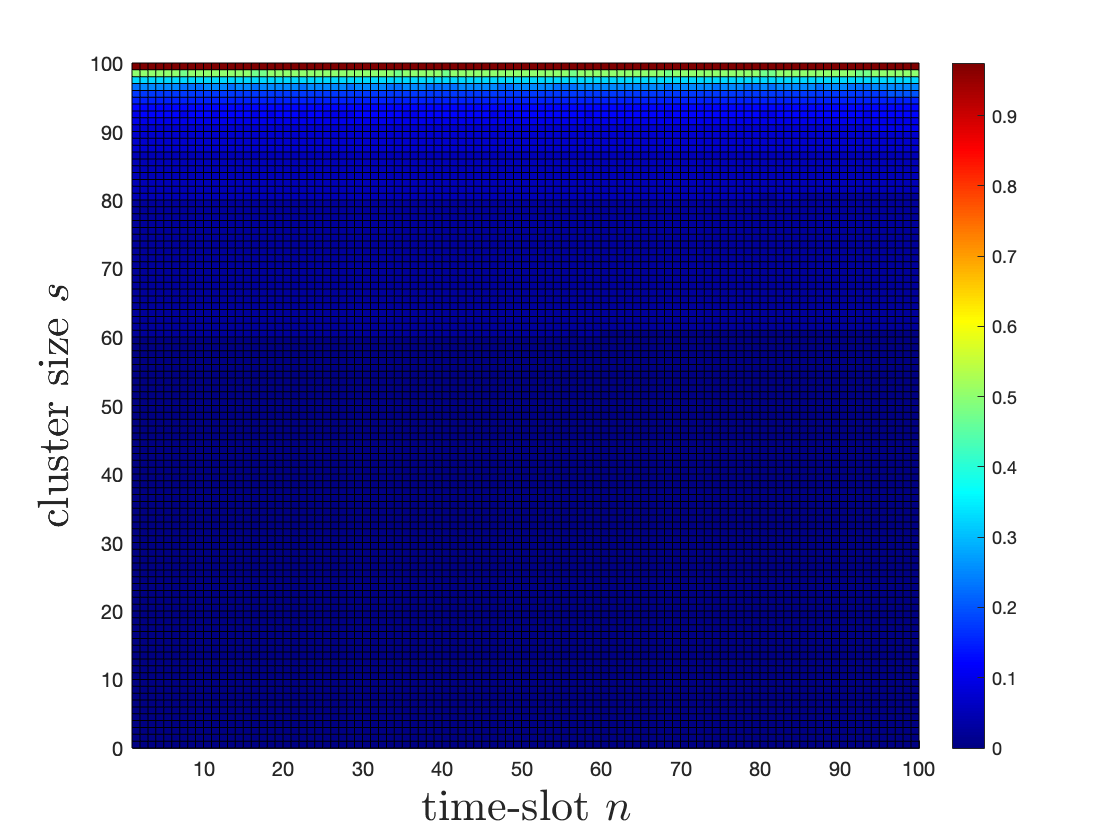}
	    \subcaption{Action matrix for pay-off function $g(s_n) = \frac{s}{S} - \frac{n}{N}$.}
	   \label{fig:09-c}
    \end{subfigure}
    \caption{Action matrices: compact representation of the optimal policy $\pi^*(s_n)$ as a function of the system state $s_n = (s,n)$ and ebit distribution probability $p$. Setting: $S=100$ and $N=100$.}
    \label{fig:09}
    \hrulefill
\end{figure*}

Finally, with the latest experiment, we aim at discussing the impact of the rewards settings  -- and, specifically, of the three base-cases introduced in \eqref{eq:30}-\eqref{eq:32} -- on the overall entanglement distribution process.

For this, we preliminary compare the optimal policy $\pi^*$ for the different settings of the pay-off function via the action matrices represented in Fig~\ref{fig:09}. Formally, the action matrix $A^*: S \times N \longrightarrow p \in [0,1]$ is defined as follows:
\begin{equation}
    a_{s,n}^* \in A^* = \tilde{p} \; \Longleftrightarrow \; \pi^*(s_n) = \begin{cases}
        Q & \forall p \leq \tilde{p} \\
        C & \forall p > \tilde{p}
    \end{cases}
\end{equation}
As an example, by considering the action map for the pay-off function $g(s_n) = \frac{s}{n}$ represented in Fig.~\ref{fig:09-a}, we note that, for an arbitrary time-slot $n$, $a_{s,n}^*$ increases as the cluster size $s$ increases. This means that, as the cluster size $s$ increases, higher values of $p$ are needed for having action $C$ being the optimal action. Clearly, for a given $n$,
for the lowest values of $s$, action $C$ is optimal for almost all the values of $p$. This is very reasonable: when the current cluster size $s$ is very small, so is the pay-off reward. Hence, it is likely more convenient to attempt another entanglement distribution rather than to stop here. And, vice-versa, for the highest values of $s$, action $Q$ is optimal for almost all the values of $p$. 

\noindent Furthermore, we observe that the values of the action matrices in Fig.~\ref{fig:09} strongly depend on the particular pay-off function.

\begin{figure*}[t]
    \begin{minipage}[t] {0.49\textwidth}
        \centering
        \includegraphics[width=\columnwidth]{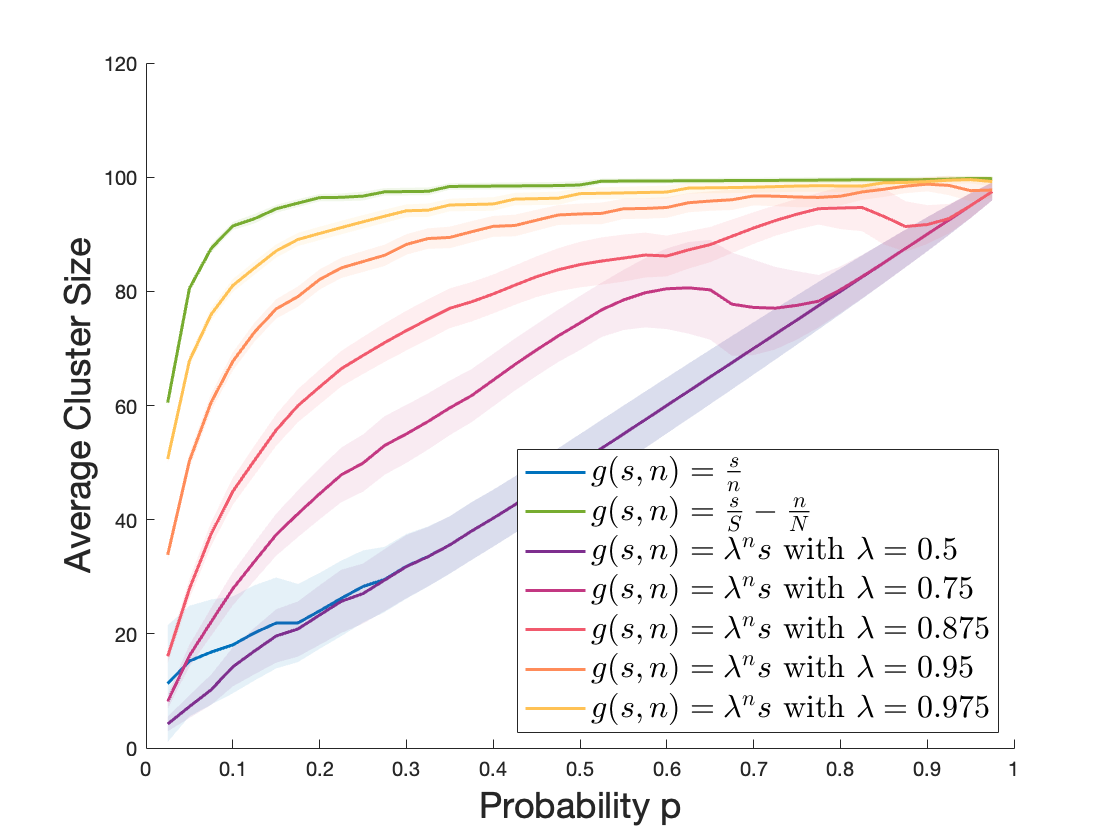}
	    \caption{Average cluster size as a function of the ebit propagation probability $p$ for $S=100$, $N=100$ and different settings of the pay-off function $g(\cdot)$. Lines denote the mean, whereas shading areas denote the standard deviation.}
	    \label{fig:07}
    \end{minipage}
    \hspace{0.02\textwidth}
    \begin{minipage}[t]{0.49\textwidth}
        \centering
        \includegraphics[width=\columnwidth]{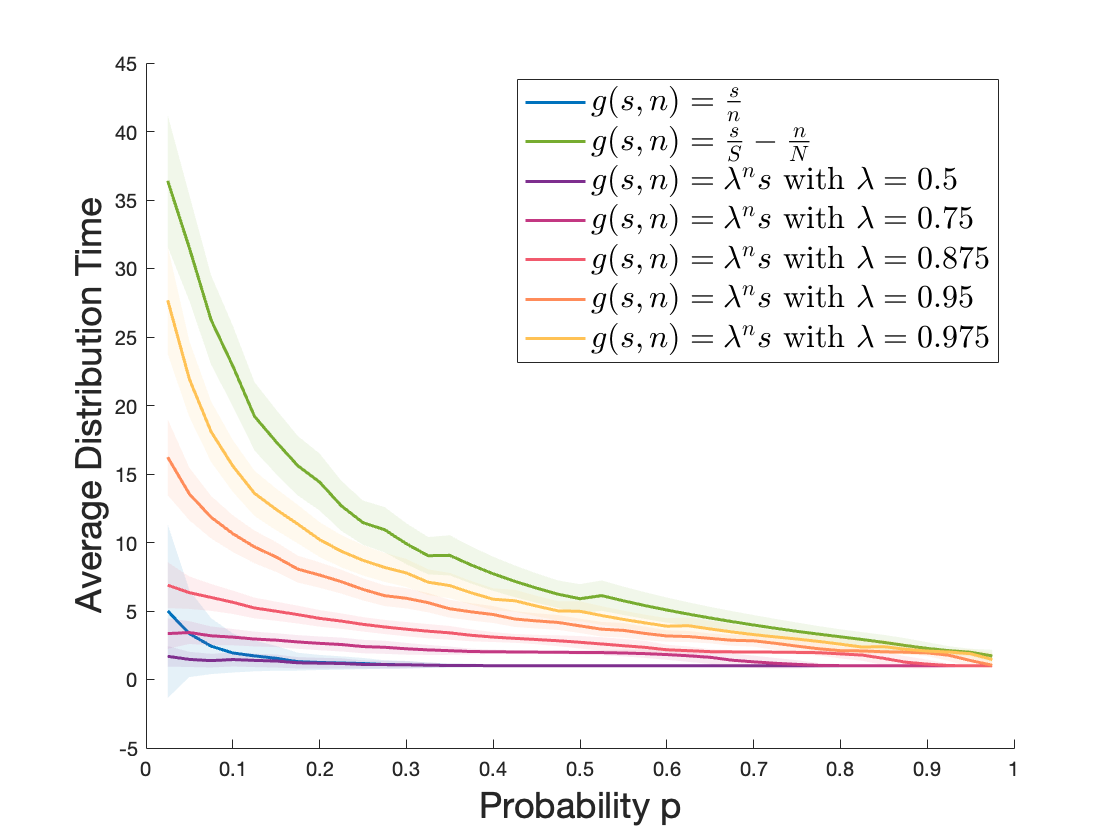}
	    \caption{Average distribution time as a function of the ebit propagation probability $p$ for $S=100$, $N=100$ and different settings of the pay-off function $g(\cdot)$. Lines denote the mean, whereas shading areas denote the standard deviation.}

	\label{fig:08}
    \end{minipage}
    \hrulefill
\end{figure*}

As instance, the action map for the pay-off function $g(s_n) = \lambda^{n}$ represented in Fig.~\ref{fig:09-b} strongly depends on the cluster size, whereas it is largely independent from the time-slot. As a result, the pay-off function $g(s_n) = \lambda^{n}$ drives the entanglement distribution process towards larger cluster sizes at the price of significantly longer distribution times.

\noindent These considerations are are clearly confirmed by Fig.~\ref{fig:07}, which presents the average cluster size $s$ as a function of the ebit propagation probability $p$ -- for the same $10^6$ trials of Fig.~\ref{fig:04} -- for the different settings of the pay-off function given in \eqref{eq:30}-\eqref{eq:32}. As before, lines denote the mean over the different trials, whereas shading areas denote the standard deviation of the different trials.

\noindent First, we note that the larger is the parameter $\lambda$ in \eqref{eq:31}, the larger is the average cluster size $s$ and the steeper is the slope of the related curve. As a matter of fact, the largest values of the average cluster size are achieved when the pay-off function is $g(s_n) = \frac{s}{S}-\frac{n}{N}$ as in \eqref{eq:32}. This agrees with the action matrix in Fig~\ref{fig:09-c}, where action $Q$ becomes optimal only for the largest values of $s$.

\noindent Interestingly, the pay-off functions significantly impact the performances for lower values of $p$. Indeed, both in Fig.~\ref{fig:07} and Fig.~\ref{fig:08}, as $p$ increases, the distance between the curves in the graph tends to reduce. The rationale is that, as $p$ increases, the target system state -- namely, the system state maximizing the reward -- can be quickly achieved in shorter distribution times.
Thus, different reward functions result in vastly different ebit distribution performances under bad transmission conditions.

\begin{remark*}
From the above, it becomes evident that, whenever there exist requirements in terms of average cluster size or average distribution time, our modeling allows to meet the performance requirements by choosing a suitable reward function, as instance by tuning the value of $\lambda$ in $g(s_n) = \lambda^n s$. Thus, our formulation of the entanglement distribution process as an optimal decision problem constitutes an effective, handy tool for quantum network designers aiming at engineering the entanglement distribution process.    
\end{remark*}


\section{Discussion and Conclusion}
\label{sec:5}

There is a wide consensus on the key role played by multipartite entanglement for quantum network functionalities and for enabling several applications in the Quantum Internet\cite{IllCalMan-22, PirDur-19,CacIllCal-23,BenHajVan-23,RauBri-01,DurBri-07}.
As instance, the unconventional features of multipartite entanglement enable a wider concept of network connectivity referred to as \textit{on-demand} connectivity\cite{IllCalMan-22}. In a nutshell, multipartite entangled states, such as GHZ states, enable the extraction of an EPR pair, which acts as point-to-point unicast link, as instance by allowing the transmission of a qubit through teleportation. Such a link can be extracted between any couple of nodes sharing the multipartite state through local operations and classical communications. Remarkably, the identities of the nodes sharing the extracted pair can be selected \textit{on-demand}, i.e., at run-time and according to the communication needs. In other words, it is possible to select the couple of nodes exploiting the extracted EPR pair afterward the entanglement generation and distribution and according to the actual communication needs. Additionally, multipartite entanglement is able to fulfill a crucial network functionality such as the \textit{Entanglement Access Control}\cite{IllCalVis-23}. To elaborate more, entanglement represents a resource for which the network nodes compete. Surprisingly, multipartite entanglement enables a collision-free strategy able to solve the contention of the entangled resource. Furthermore and as recently discussed in \cite{BenHajVan-23}, multipartite entangled states -- such as graph states -- enable an all-photonic repeater scheme for quantum repeater networks which promises tolerance to photon losses. Besides, multipartite entangled states represent a computing resource. As instance, entangled states such as cluster states are widely known for their ability to implement the so-called \textit{measurement-based quantum computing} model \cite{RauBri-01} and to enable quantum error correction strategies \cite{DurBri-07}.

With the above discussion in mind, it comes straightforward that multipartite entanglement is a key resource for various quantum network functionalities and applications, each presenting distinct requirements and serving different purposes. As a consequence, multipartite entanglement distribution demands careful engineering. To this aim, a crucial step involves the formulation of a general model capable of not only assessing the impact of noise on the distribution process but also positioning itself as a versatile tool for the diverse applications of multipartite entanglement.

This work moves a step towards the aforementioned direction, by developing 
an handy-tool for tuning and engineering the entanglement distribution process so that it can meet the performance requirements through proper reward functions. The developed theoretical framework jointly accounts for the constraints arising from the underlying technologies as well as for the overlaying communication protocol requirements.  We exploited our formulation for discussing the trade-off arising between the two key performance metrics -- i.e., the average cluster size and the average distribution time -- and for discussing the impact of the reward function and the decision-making policy on the entanglement distribution performance.

\begin{appendices}

\section{Proof of Lemma~\ref{lem:01}}
\label{app:01}

According to the model developed in Sec.~\ref{sec:2}, a distribution attempt takes place only if action $a=C$ is taken. And in this case, at time slot $n+1$, the system -- as a result of the distribution attempts -- evolves into another state characterized by a number $\tilde{s}$ of ``connected nodes'', which cannot be smaller than the number $s$ of ``connected nodes'' in the time slot $n$. The reason for which $\tilde{s} \geq s$ is twofold: i) the heralded scheme allows the super-node to recognize which node – if any – experienced an ebit loss in a given time-slot. Hence, in the successive time slot, the super-node distributes entanglement only to the missing nodes; ii) by restricting the distribution attempts within a time interval $N$ where the decoherence effects are negligible, the system state evolution is restricted from “backward” transitions towards smaller connected sets with $\tilde{s} < s$.
Stemming from this and in according to the EPR distribution model given in Sec.~\ref{sec:2}, each ebit distribution attempt follows a Bernoulli distribution with parameter $p$. Accordingly, it follows that when $a = C$ and $s,\tilde{s} \in \mathcal{S} : \tilde{s} \geq s$, the transition probability $p(\tilde{s}_{n+1}|s_n,C)$ is given by:
\begin{equation}
        \label{eq:app_01} 
        p(\tilde{s}_{n+1}|s_n,C) = 
            p(\tilde{s}|s)= \binom{S-s}{\tilde{s}-s}q^{S-\tilde{s}}p^{\tilde{s}-s}.
\end{equation}

\noindent Conversely, when action $a=Q$ is taken, the system can only evolve in the absorption state $\Delta$, which is a fictitious state modeling the state where no further distribution attempts are performed. It is worthwhile to observe that once in the absorption state $s=\Delta$, the system remains in such a state, i.e., no evolution towards $\tilde{s} \neq \Delta $ is allowed. As a consequence the proof follows.

\section{Proof of Lemma~\ref{lem:02}}
\label{app:02}
We have two statements to prove within the inequality given in \eqref{eq:25}.
\begin{IEEEproof}
    \textit{First Inequality}.
    
    We start by proving the first part of the inequality in \eqref{eq:25}, namely:
    \begin{equation}
        \label{eq:app.1}
        v^-(s_n) \leq v^*_C(s_n) \; \forall \, s \in \mathcal{S} \wedge n < N
    \end{equation}
    By exploiting the expression of $v^*_C(s_n)$ in \eqref{eq:20}, one can recognize that:
    \begin{equation}
        \label{eq:app.2}
        v^*_C(s_n)= -f(s_n) + \sum_{\tilde{s} \in \tilde{\mathcal{S}}} p\big( \tilde{s}_{n+1} | s_{n} , C \big) v^*\big(\tilde{s}_{n+1} \big)
    \end{equation}
    According to \eqref{eq:21}, $v^*\big(\tilde{s}_{n+1} \big)\geq v_Q^*(\tilde{s}_n+1)$ and by accounting for the expression of $v^-(s_n)$ in \eqref{eq:24}, the proof follows.
\end{IEEEproof}

\begin{IEEEproof}
    \textit{Second Inequality}.
    
    We now prove the second part of the inequality in \eqref{eq:25}, i.e.:
    \begin{equation}
        \label{eq:app.3}
        v^+(s_n) \geq v^*_C(s_n)\; \forall \, s \in \mathcal{S} \wedge n < N
    \end{equation}
    By exploiting the expressions of $v^+(s_n)$ in \eqref{eq:23} and $v_C^*(s_n)$ reported in \eqref{eq:app.2}, one recognizes that proving \eqref{eq:app.3} is equivalent to prove that:
    \begin{equation}
        \label{eq:app.4}
         \sum_{\breve{s} \in \tilde{\mathcal{S}}} p\big( \breve{s}_{N} | s_{n} , C \big) v_Q^*\big(\breve{s}_{n+1} \big) \geq \sum_{\tilde{s} \in \tilde{\mathcal{S}}} p\big( \tilde{s}_{n+1} | s_{n} , C \big) v^*\big(\tilde{s}_{n+1} \big),
    \end{equation}
    where, by definition
    \begin{equation}
        \label{eq:app.5}
         v^*\big(\tilde{s}_{n+1} \big) = \max \left\{  v^*_Q(\tilde{s}_{n+1}), v^*_C(\tilde{s}_{n+1}) \right\}
    \end{equation}
    To prove \eqref{eq:app.4}, we can consider the two elements in \eqref{eq:app.5} separately. To this aim, let us consider the more general case, namely, the case where $n+1<N$\footnote{Indeed, when $n+1=N$, no decision has to be made since the distribution is interrupted and the system goes in the absorption state}. 
    
    \textbf{Case $1$: $v^*\big(\tilde{s}_{n+1} \big) = v^*_Q(\tilde{s}_{n+1})$.} \\
    Let us conduct a proof with a \textit{reductio ad absurdum}, i.e., let us suppose that:
    \begin{equation}
        \label{eq:app.6}
         \sum_{\breve{s} \in \tilde{\mathcal{S}}} p\big( \breve{s}_{N} | s_{n} , C \big) v_Q^*\big(\breve{s}_{n+1} \big) < \sum_{\tilde{s} \in \tilde{\mathcal{S}}} p\big( \tilde{s}_{n+1} | s_{n} , C \big) v^*_Q(\tilde{s}_{n+1}) \big).
    \end{equation}
    By accounting for the extended transition probabilities given in \eqref{eq:22}, we obtain equation \eqref{eq:app.7} given at the top of the next page.
    \begin{figure*}
    \begin{equation}
        \label{eq:app.7}
         \sum_{\breve{s} \geq s} p\big( \breve{s}_{N} | s_{n} , C \big) g(\breve{s}_{n+1}) =
            \sum_{\breve{s} \geq \tilde{s}} \sum_{\tilde{s} \geq s}^{\breve{s}} p\big( \breve{s}_{N} | \tilde{s}_{n+1} , C \big) p\big( \tilde{s}_{n+1} | s_{n} , C \big) g(\breve{s}_{n+1} )
            < \sum_{\tilde{s} \geq s} p\big( \tilde{s}_{n+1} | s_{n} , C \big) g(\tilde{s}_{n+1}) 
    \end{equation}
    \hrulefill
    \end{figure*}
    We note that \eqref{eq:app.7} is satisfied only if there exists at least one $\tilde{s} \in \mathcal{S}: \tilde{s} \geq s$ so that:
    \begin{align}
        \label{eq:app.8}
            & \sum_{\breve{s} \geq \tilde{s}} p\big( \breve{s}_{N} | \tilde{s}_{n+1} , C \big) p\big( \tilde{s}_{n+1} | s_{n} , C \big) g(\breve{s}_{n+1} ) < \nonumber \\
            & \quad \quad \quad < p\big( \tilde{s}_{n+1} | s_{n} , C \big) g(\tilde{s}_{n+1}) \Longleftrightarrow \nonumber \\
            & \Longleftrightarrow \sum_{\breve{s} \geq \tilde{s}} p\big( \breve{s}_{N} | \tilde{s}_{n+1} , C \big) g(\breve{s}_{n+1} ) < g(\tilde{s}_{n+1}).
    \end{align}
    By accounting for Property~\ref{prop:01} and by recognizing that $\sum_{\breve{s} \geq \tilde{s}} p\big( \breve{s}_{N} | \tilde{s}_{n+1} , C \big) = 1$, \eqref{eq:app.8} constitutes a \textit{reductio ab absurdum} and so does \eqref{eq:app.6}.

    \textbf{Case $2$: $v^*\big(\tilde{s}_{n+1} \big) = v^*_C(\tilde{s}_{n+1})$.} \\
    Let us conduct the proof again with a \textit{reductio ad absurdum} by supposing that:
    \begin{equation}
        \label{eq:app.9}
         \sum_{\breve{s} \in \tilde{\mathcal{S}}} p\big( \breve{s}_{N} | s_{n} , C \big) v_Q^*\big(\breve{s}_{n+1} \big) < \sum_{\tilde{s} \in \tilde{\mathcal{S}}} p\big( \tilde{s}_{n+1} | s_{n} , C \big) v^*_C(\tilde{s}_{n+1}) \big)
    \end{equation}
    By accounting for the extended probabilities given in \eqref{eq:22}, we obtain equation \eqref{eq:app.10} given at the top of the next page.
    \begin{figure*}
        \begin{equation}
            \label{eq:app.10}
            \sum_{\breve{s} \geq \tilde{s}} \sum_{\tilde{s} \geq s}^{\breve{s}} p\big( \breve{s}_{N} | \tilde{s}_{n+1} , C \big) p\big( \tilde{s}_{n+1} | s_{n} , C \big) g(\breve{s}_{n+1} )
                < \sum_{\tilde{s} \geq s} p\big( \tilde{s}_{n+1} | s_{n} , C \big) \left( -f(\tilde{s}_{n+1}) + \sum_{\breve{s} \geq \tilde{s}} p\big( \breve{s}_{n+2} | \tilde{s}_{n+1} , C \big) v^*(\breve{s}_{n+2}) \right)
        \end{equation}
        \hrulefill
    \end{figure*}
    For the sake of notation simplicity and with no loss in generality -- as discussed at the end of this proof -- let us assume $N=n+2$. Accordingly, $v^*(\breve{s}_{n+2}) = g(\breve{s}_{N})$ and \eqref{eq:app.10} holds only if there exists at least one $\tilde{s} \in \mathcal{S}: \tilde{s}\geq s$ so that:
    \begin{align}
        \label{eq:app.11}
         & \sum_{\breve{s} \geq \tilde{s}} p\big( \breve{s}_{N} | \tilde{s}_{n+1} , C \big) g(\breve{s}_{n+1} )
             < \nonumber \\
             & \quad < - f(\tilde{s}_{n+1}) + \sum_{\breve{s} \geq \tilde{s}} p\big( \breve{s}_{N} | \tilde{s}_{n+1} , C \big) g(\breve{s}_{N})
    \end{align}
    Hence, by accounting for Property~\ref{prop:02}, \eqref{eq:app.11} constitutes a \textit{reductio ab absurdum} and so does \eqref{eq:app.9}. We finally note that, whether $N$ should be greater than $n+2$ -- say $N = n+3$ as instance -- we have that $v^*(\breve{s}_{n+2})$ is equal to $\max \left\{  v^*_Q(\tilde{s}_{N-1}), v^*_C(\tilde{s}_{N-1}) \right\}$, and the proof follows recursively by adopting the same reasoning adopted for the two elements in \eqref{eq:app.5}.
\end{IEEEproof}

\section{Proof of Proposition~\ref{prop:02}}
\label{app:03}

\subsection{Case I: rewards modeled as in \eqref{eq:30}.}

Here we prove that the OLA rule is not optimal when the rewards are modeled as in \eqref{eq:30}, namely, when:
\begin{equation}
    \label{eq:48}
    g(s_n) = \frac{s}{n}
\end{equation}
Let us assume the system state being $s_n \in \mathcal{S}$. Whether action $C$ is chosen, the expected state $E[\mathfrak{S}_{n+1}]$ is given by:
\begin{equation}
    \label{eq:49}
    E[\mathfrak{S}_{n+1}] = \sum_{\tilde{s} \in \mathcal{S}} \binom{S-s}{\tilde{s} - s} q^{S-\tilde{s}} p^{\tilde{s}-s} = s + p (S-s)
\end{equation}
Accordingly, stemming from the definition of OLA set in \eqref{eq:27} and by accounting for \eqref{eq:29}, we have that $S^Q_n$ and $S^Q_{n+1}$ are given by:
\begin{align}
    \label{eq:50}
    & S^Q_n = \left\{ x \in \mathcal{S} : \frac{x}{n} \geq \frac{x + p (S-x)}{n+1} \right\} \\
    \label{eq:51}
    & S^Q_{n+1} = \left\{ x \in \mathcal{S} : \frac{x}{n+1} \geq \frac{x + p (S-x)}{n+2} \right\}
\end{align}
Hence, after simple algebraic manipulations, it results:
\begin{align}
    \label{eq:52}
    & s \in S^Q_n \Longrightarrow s \geq \frac{n p}{1 + n p} S \\
    \label{eq:53}
    & \tilde{s} \in S^Q_{n+1} \Longrightarrow \tilde{s} \geq \frac{(n+1) p}{1 + (n+1) p} S
\end{align}
Let us conduct the proof with a \textit{reductio ab absurdum} by assuming that, starting from state $s_n : s \in S^Q_n$ and evolving into state $\tilde{s}_{n+1}$, it must result $\tilde{s} \in S^Q_{n+1}$ for any $\tilde{s}$. Without any loss of generality, we assume:
\begin{equation}
    \label{eq:54}
    s = \frac{n p}{1 + n p} S \quad \wedge \quad \tilde{s} = s
\end{equation}
and, by jointly accounting for \eqref{eq:53} and \eqref{eq:54}, it results:
\begin{equation}
    \label{eq:55}
    \tilde{s} = s = \frac{n p}{1 + n p} S > \frac{(n+1) p}{1 + (n+1) p} S \Longrightarrow p < 0
\end{equation}
which clearly constitutes a \textit{reductio ab absurdum}.

\subsection{Case II: rewards modeled as in \eqref{eq:31}.}

Here we prove that the OLA rule is optimal when the rewards are modeled as in \eqref{eq:31}, namely, when:
\begin{equation}
    \label{eq:56}
    g(s_n) = \lambda^n s
\end{equation}
To this aim, let us assume $s_n \in \mathcal{S}^{Q}_n$ and let us conduct the proof with a \textit{reductio ab absurdum} by assuming that the system can evolve into a $\tilde{s}_{n+1} \notin \mathcal{S}^{Q}_{n+1}$. From \eqref{eq:49}, we have that $S^Q_n$ and $S^Q_{n+1}$ are given by:
\begin{align}
    \label{eq:57}
    &S^Q_n = \left\{ x \in \mathcal{S} : \lambda^n x \geq \lambda^{n+1}  x + p (S-x) \right\} \\
    \label{eq:58}
    &S^Q_{n+1} = \left\{ x \in \mathcal{S} : \lambda^{n+1} x \geq \lambda^{n+2}  x + p (S-x) \right\}
\end{align}
Hence, after simple algebraic manipulations, it results:
\begin{align}
    \label{eq:59}
    & s \in S^Q_n \Longrightarrow s \geq \frac{\lambda p S}{1 - \lambda - \lambda p} \\
    \label{eq:60}
    & \tilde{s} \notin S^Q_{n+1} \Longrightarrow \tilde{s} < \frac{\lambda p S}{1 - \lambda - \lambda p}
\end{align}
which constitutes a \textit{reductio ab absurdum}, given that the system cannot evolve from $s_n$ to $\tilde{s}_{n+1}$ with $\tilde{s} < s$.

\subsection{Case III: rewards modeled as in \eqref{eq:32}.}

Here we prove that the OLA rule is optimal when the rewards are modeled as in \eqref{eq:32}, namely, when:
\begin{equation}
    \label{eq:61}
    g(s_n) = \frac{s}{S} - \frac{n}{N}
\end{equation}
To this aim, let us assume $s_n \in \mathcal{S}^{Q}_n$ and let us conduct the proof with a \textit{reductio ab absurdum} by assuming that the system can evolve into a $\tilde{s}_{n+1} \notin \mathcal{S}^{Q}_{n+1}$. From \eqref{eq:49}, we have that $S^Q_n$ and $S^Q_{n+1}$ are given by:
\begin{align}
    &S^Q_n = \left\{ x \in \mathcal{S} : \frac{x}{S} - \frac{n}{N} \geq \frac{x + p x}{S} + p + \frac{n+1}{N} \right\} \\
    &S^Q_{n+1} = \left\{ x \in \mathcal{S} : \frac{x}{S} - \frac{n+1}{N} \geq \frac{x + p x}{S} + p + \frac{n+2}{N} \right\}
\end{align}
Hence, after simple algebraic manipulations, it results:
\begin{align}
    & s \in S^Q_n \Longrightarrow s \geq S - \frac{S}{N p} \\
    & \tilde{s} \notin S^Q_{n+1} \Longrightarrow \tilde{s} < S - \frac{S}{N p}
\end{align}
which constitutes a \textit{reductio ab absurdum}, given that the system cannot evolve from $s_n$ to $\tilde{s}_{n+1}$ with $\tilde{s} < s$.

\end{appendices}

\bibliographystyle{IEEEtran}
\bibliography{biblio.bib}

\end{document}